\documentclass[%
reprint,superscriptaddress,
 amsmath,amssymb,
 aps,
floatfix,
]{revtex4-2}
\UseRawInputEncoding
\usepackage{graphicx}
\usepackage{dcolumn}
\usepackage{bm}
\usepackage{hyperref}
\usepackage{xcolor}
\usepackage[caption = false]{subfig}
\usepackage[utf8]{}
\usepackage{float}
\usepackage{amsmath}
\usepackage{amsfonts}
\DeclareMathOperator\erf{erf}

\begin{document}

\title{Lotka--Volterra predator-prey model with periodically varying carrying capacity}

\author{Mohamed Swailem} \email{mswailem@vt.edu}
\affiliation{Department of Physics \& Center for Soft Matter and Biological Physics, MC 0435, Robeson Hall, 
		850 West Campus Drive, Virginia Tech, Blacksburg, VA 24061, USA}
\author{Uwe C. T{\"a}uber} \email{tauber@vt.edu}
\affiliation{Department of Physics \& Center for Soft Matter and Biological Physics, MC 0435, Robeson Hall, 
		850 West Campus Drive, Virginia Tech, Blacksburg, VA 24061, USA}
\affiliation{Faculty of Health Sciences, Virginia Tech, Blacksburg, VA 24061, USA}

\date{\today}

\begin{abstract}
We study the stochastic spatial Lotka--Volterra model for predator-prey interaction subject 
to a periodically varying carrying capacity. 
The Lotka--Volterra model with on-site lattice occupation restrictions (i.e., finite local carrying capacity) that
represent finite food resources for the prey population exhibits a continuous active-to-absorbing phase 
transition.
The active phase is sustained by the existence of spatio-temporal patterns in the form of 
pursuit and evasion waves.
Monte Carlo simulations on a two-dimensional lattice are utilized to investigate the effect of seasonal 
variations of the environment on species coexistence. 
The results of our simulations are also compared to a mean-field analysis in order to specifically delineate
the impact of stochastic fluctuations and spatial correlations. 
We find that the parameter region of predator and prey coexistence is enlarged relative to 
the stationary situation when the carrying capacity varies periodically.
The (quasi-)stationary regime of our periodically varying Lotka--Volterra predator-prey system shows 
qualitative agreement between the stochastic model and the mean-field approximation.
However, under periodic carrying capacity switching environments, the mean-field rate equations predict 
period-doubling scenarios that are washed out by internal reaction noise in the stochastic lattice model. 
Utilizing visual representations of the lattice simulations and dynamical correlation functions, we study how 
the pursuit and evasion waves are affected by ensuing resonance effects.
Correlation function measurements indicate a time delay in the response of the system to sudden changes in 
the environment. 
Resonance features are observed in our simulations that cause prolonged persistent spatial correlations.
Different effective static environments are explored in the extreme limits of fast- and slow periodic switching.
The analysis of the mean-field equations in the fast-switching regime enables a semi-quantitative description 
of the (quasi-)stationary state.
\end{abstract}

\maketitle

\section{\label{section:introduction}Introduction\protect\\}

The study of population dynamics has gained popularity among various fields of research in recent years 
\cite{math1,math2,math3,math4,math5,math6,math7,math8,math9,math10,math11,waves1,physics1,
physics2,physics3,physics4,physics5,physics6,physics8,physics9, uwe4, uwe00,envxx1,envxx2}. 
Ecosystems of multiple interacting species are traditionally modelled as a dynamical system described by a 
set of deterministic differential rate equations. 
Yet such deterministic descriptions do not capture the stochastic nature of real-life systems and ignore 
temporal and spatial correlation effects that certainly affect the system's quantitative features, and maybe
its qualitative behavior \cite{rick1,rick2}. 
Therefore, various efforts have been made in trying to adequately represent such systems in terms of 
coupled stochastic processes \cite{math9,physics1, physics2, physics5, physics8, doi}. 
An additional difficulty in the study of stochastic population dynamics stems not just from the fact that they 
are non-linear dynamical systems with a large number of degrees of freedom, but also because they do not 
reside in thermal equilibrium: 
Hence the stationary probability distribution is not the standard Boltzmann distribution, non-vanishing 
probability currents decisively characterize the ensuing non-equilibrium steady states, and irreversibility is
crucial, as becomes manifest in absorbing states that characterize population extinction. 
However, lattice simulations can be used effectively to gain insight on the interacting populations' behavior, 
and may thus guide the development of new techniques for studying non-equilibrium systems. 

This study focuses on the paradigmatic predator-prey model introduced independently by Lotka and 
Volterra \cite{lotka,volterra}, owing to its simplicity and extensive prevalent literature. 
The original formulation of the Lotka--Volterra model utilized a coupled set of deterministic differential 
equations describing the temporal evolution of the predator and prey densities. 
It was successful in explaining population oscillations that are present in predator-prey ecologies. 
However, the Lotka--Volterra mean-field model was aptly met with criticism because it did not account for 
stochastic fluctuations, and since it predicts stable density oscillations that are fully determined by the
initial population densities, whereas in nature, predator-prey systems can exhibit extinction or fixation. 
The neutral limit cycles of the original Lotka--Volterra model are also not stable under straightforward 
modifications to the model \cite{math2,math4, uwe4}:
Allowing for intrinsic stochastic noise, or introducing a finite carrying capacity render the limit cycles 
unstable, and the system is instead driven to a stable fixed point with constant predator and prey densities. 
We remark that there exist alternative predator-prey models that can predict stable limit 
cycles such as those discussed in Refs.~\cite{RM-Model1, RM-Model2}. 
Further it is well-established that spatial structure in ecological systems promotes species coexistence 
\cite{spatial1, spatial2, spatial3, spatial4}. 
This asserrtion was supported by experiments done by Huffaker et al. \cite{experiment2}, who found that 
coexistence of a predator-prey system of mite species was maintained via spatial heterogeneity of species 
densities. 
This was later hypothesized to be a result of asynchronous system states in different patches (lattice sites) \cite{async1,spatial2,spatial4}.

A substantial body of experimental work has been performed on ecologies that exhibit predator-prey type 
interactions \cite{math10, experiment1, experiment2, experiment3, experiment4, 
experiment5, experiment-ref2}.
While the Lotka--Volterra model is able to capture the periodic behavior of such systems, with good numerical 
agreement for well-mixed microbial systems \cite{experiment5, experiment1}, its mean-field approximation 
cannot capture the stochastic fluctuations in the population densities. 
As pointed out in Ref.~\cite{experiment1}, the issue is that the deterministic Lotka--Volterra model (even with 
a finite carrying capacity) only allows for decaying or constant oscillation amplitudes.
It is hence preferable to consider the Lotka--Volterra model as a stochastic reaction-diffusion system 
incorporating the following reactions that involve the predator species $A$ and the prey species $B$:
\begin{subequations}
	\label{eq:reactions}
	\begin{eqnarray}
	&A \xrightarrow{\mu} \emptyset, \quad &\text{predator death,} \label{subeq:pred_death} \\
	&B \xrightarrow{\sigma} B+B, \quad &\text{prey reproduction (birth),} \ \label{subeq:prey_birth} \\
	&A+B \xrightarrow{\lambda} A+A, \quad &\text{predation,} \label{subeq:predation}
	\end{eqnarray}
\end{subequations}
where $\mu$, $\sigma$, and $\lambda$ denote the corresponding reaction rates that quantitatively characterize 
the stochastic processes. 
The reaction (\ref{subeq:predation}) combines the actions of simultaneous predation and 
predator reproduction, a common simplification \cite{math5, physics1, lattice2, lattice3, lattice4, critical2, 
physics2, physics3, physics6, vary_env0, vary_env2, vary_env3, uwe1, uwe2, uwe3, uwe4}; as shown in
Ref.~\cite{uwe00}, for spatially extended stochastic realizations of the Lotka--Volterra processes, separating  
(\ref{subeq:predation}) into two independent reactions does not qualitatively change the stochastic, spatially 
extended system's behavior.

This simplest Lotka--Volterra model variant can be readily extended to account for finite resources for the prey 
population.
On the mean-field level, one may just add a logistic growth limiting factor for the prey species
\cite{logistic1,logistic2,math4}.
For the stochastic model realized on a regular lattice, this can be achieved by implementing on-site lattice 
occupation restrictions \cite{lattice1,lattice2,lattice3, lattice4, lattice5,physics1,physics2,physics5,physics8,uwe1}. 
An alternative method of modelling competition between prey individuals for resources would be to implement 
the binary reaction $B+B \xrightarrow{}B$, which provide a ``soft" local particle number constraint \cite{uwe3}. 
In contrast to imposing ``hard" on-site restrictions in the lattice model, the corresponding mean-field rate
equation would directly lead to a logistic equation.
Either modification of the stochastic Lotka--Volterra model induces a continuous non-equilibrium phase 
transition between two-species coexistence and predator extinction. 
If the predators are not efficient in hunting their prey, or if the food resources available to the prey are scarce, 
the predator population eventually goes extinct \cite{physics1, physics2, physics8, uwe1, uwe2, uwe3}.
The critical exponents of this active-to-absorbing state phase transition were shown to be in the directed 
percolation universality class by means of numerical simulations \cite{physics8,lattice4,lattice5,critical1,critical2, 
critical3} as well as a field-theoretic analysis \cite{uwe2,uwe3, uwe4}.
Persistent spatio-temporal structures emerging in the coexistence phase of the stochastic lattice Lotka--Volterra 
model that substantially enhance species coexistence and thus promote ecological diversity have been thoroughly 
studied as well \cite{spatial4,waves1,waves2,waves3,physics4,uwe2}.
Prominent travelling pursuit and evasion waves arise due to the fact that predators must move towards high 
concentrations of prey in order to survive, leaving behind them areas of low prey concentration, while the prey 
similarly need to evade regions of high predator densities. 
These waves lead to asynchronous states and therefore enhance coexistence \cite{spatial4}, which underscores 
the importance of spatial modelling for predator-prey systems.

Experimental in-vitro as well as in-vivo systems are often exposed to varying nutrients, which affects species 
survival. 
Therefore the modelling of population dynamics with temporally varying environments has gained attention in 
recent years \cite{volterra-ref, varyenv-ref1, varyenv-ref2, varyenv-ref3, vary_env0,vary_env1,vary_env2,
vary_env3,vary_env4,vary_env5,vary_env6,vary_env7,vary_env8,vary_env9,vary_env10,vary_env11,
vary_env12,vary_env13}.
Traditionally, fluctuations in the environment are modelled as variable reaction rates 
\cite{varyenv-ref1, vary_env1, vary_env2, vary_env3, vary_env8, vary_env9, vary_env10, vary_env11} which 
usually enter linearly. 
On the other hand, to investigate the effects of varying non-linear parameters, typically time-dependent carrying 
capacities are introduced \cite{vary_env4, vary_env5,vary_env6,vary_env7}, but in a non-spatial setting. 
Yet spatial models with a varying carrying capacity have also not been properly explored in the literature. 
Lattice models are often simulated with a fixed on-site restriction \cite{lattice1,lattice2,lattice3, lattice4, 
lattice5,physics1,physics2,physics5,physics8,uwe1}.

In this study, in order to gain a full understanding of how a time-varying on-site resource constraint can change 
the quasi-stationary properties as well as transient kinetics of predator-prey competition dynamics, we consider 
the stochastic Lotka--Volterra model on a regular two-dimensional lattice (with periodic boundary conditions) 
with a finite local prey carrying capacity that varies periodically over time. 
This oscillatory environmental variability resembles seasonal changes in food availability for the prey population. 
While seasonal changes may additionally affect other parameters such as the reproduction rate, in this study we 
focus on the effects of temporal oscillations in resource availability, since we anticipate variability in this non-linear 
parameter to generate the most prominent modifications relative to the stationary case.
This variation in the environment leads to indefinite populations oscillations, whereas the static Lotka--Volterra 
model only supports damped oscillations with a decreasing amplitude (in the coexistence regime). 
Similar conclusions were already drawn in Ref.~\cite{varyenv-ref3}.
We investigate how a sudden increase in prey food resources can prevent the predators from going extinct. 
Specifically, intriguing dynamical behavior is observed when the system switches between carrying capacity 
values that would result in species coexistence and predator extinction, respectively, in stationary environments.
One may regard this Lotka--Volterra system with periodically varying environment as a dynamical system 
subject to an oscillating external driving force. 
In periodically driven dynamical systems, there are two limiting situations that allow for 
quantitative theoretical analysis, namely the fast and slow switching regimes, for which the driving force 
oscillation period is small or large, respectively, compared to the intrinsic oscillation time scale of the system. 
In order to quantitatively analyze our model, we measure the time evolution of the population density for each 
species and their two-point correlations functions.
We demonstrate that an analysis of the coupled mean-field rate equations allows a 
semi-quantitative description of the (quasi-)stationary state of the system for rapidly varying environments.
As mentioned  in Ref.~\cite{varyenv-ref3}, density oscillations tend to have the same period as the environmental
oscillations. However, our model exhibits period doubling effects when an asymmetric environment is considered.

Our aim is to understand the mechanism for the enlargement of the region of parameter space 
that permits species coexistence in the Lotka--Volterra predator-prey model, as a consequence of the periodic
variations in the environment. 
In the fast switching regime, we delineate under which conditions the environmental variability may be captured 
through effective averaged parameters.
Direct comparisons between mean-field and the lattice model results allows us to determine quantitatively when
the analysis of approximate mean-field rate equations suffices.
We also address the question of how the externally imposed carrying capacity dynamics interacts with the 
intrinsic spatio-temporal pursuit and evasion waves characteristic of predator-prey models.
Indeed, this interplay between the spreading population waves and the changing environment causes intriguing
resonant behavior in the system.

This paper is organized as follows: 
Section~\ref{section:mean-field} gives an overview of the stationary states of the Lotka--Volterra model for 
predator-prey competition and their stability within the mean-field theory framework. 
It next describes the features found by numerically integrating the coupled rate equations for periodically 
varying carrying capacity.
We then mathematically analyze the (quasi-) stationary state of the mean-field model in both the slow- and 
fast-switching regimes.
Our implementation for our corresponding stochastic lattice model and the ensuing simulation data are 
presented in Section~\ref{section:lattice_model}, and compared with the mean-field results.
Finally, our summary and concluding remarks are provided in Section~\ref{section:conclusion}.

\section{Lotka--Volterra predator-prey competition: mean-field theory}
\label{section:mean-field}

\subsection{Constant carrying capacity: mean-field rate equations and stability analysis}
\label{subsection:stability_analysis}

Mean-field rate equations for stochastic dynamical reaction systems are approximate deterministic equations 
that aptly describe a well-mixed setup. 
Even though they neglect spatial correlations and temporal fluctuations, they are often useful to gain intuition 
on the system's expected behavior. 
In Sec.~\ref{section:lattice_model}, we shall compare the results obtained with the mean-field equations 
with the Monte Carlo simulation data from the full stochastic model (\ref{eq:reactions}).

For the Lotka--Volterra predator-prey competition model (\ref{eq:reactions}), the classical 
mean-field rate equations that describe the time evolution of the mean predator and prey densities $a(t)$ and
$b(t)$ read
\begin{subequations}
\label{eq:mean-field0}
   \begin{eqnarray}
   \frac{da(t)}{dt} &=& -\mu a(t) +\lambda a(t) b(t) \ , \label{subeq:pred_MF0} \\
   \frac{db(t)}{dt} &=& \sigma b(t) -\lambda a(t) b(t) \ .
   \label{subeq:prey_MF0}
   \end{eqnarray}
\end{subequations}
These rate equations can be understood as representing gain / loss terms for reactions that 
increase / decrease the population densities. 
Linear stability analysis of this system shows that the system exhibits a species coexistence fixed point, and 
numerical integration of these equations leads to oscillatory behavior, namely neutral limit cycles. 
We will perform our analysis on the more generalized Lotka--Volterra model with a growth-limiting factor
for the prey species (for reviews, see Refs.~\cite{uwe2, math5}).

The original Lotka--Volterra rate equations can be generalized by including a growth-limiting 
factor $1 - a(t)/K_1-b(t) / K_2$, where $K_1$ and $K_2$ respectively represent the (global) carrying capacities 
induced by prey-predator and prey-prey resource competition. 
For simplicity, we set $K_1=K_2$, since this does not change the qualitative behavior of the system on the 
mean-field level; this implies the modified set of rate equations
\begin{subequations}
\label{eq:mean-field}
   \begin{eqnarray}
   \frac{da(t)}{dt} &=& -\mu a(t) +\lambda a(t) b(t) \ , \label{subeq:pred_MF} \\
   \frac{db(t)}{dt} &=& \sigma b(t) \left( 1-\frac{a(t)+b(t)}{K} \right) -\lambda a(t) b(t) \ , 
   \label{subeq:prey_MF}
   \end{eqnarray}
\end{subequations}
where $K$ denotes the (global) carrying capacity.
Their mean-field character resides in the assumed factorization for the non-linear predation reaction with 
rate $\lambda$ of a two-point correlation function into a mere density product, which assumes statistical
independence and the absence of correlations. 
The growth-limiting factor is used to model limited finite resources, and vanishes if $a(t)+b(t)=K$.
In that case, the prey density's temporal derivative becomes negative, indicating a strictly decreasing prey 
population. 
We remark that adding an explicit growth limiting term for the predator density is not required since the 
predators' growth is determined by the prey density. 
Hence, if the prey species has a growth limiting factor, this will indirectly constrain the predator population
abundance as well.

The stationary states of this system are given by constant solutions to (\ref{eq:mean-field}). 
This results in three fixed points $(a^*,b^*)=\{(0,0), (0,K), (a_0,b_0)\}$, where
\begin{equation}
\label{eq:stationary}
    a_0 = \frac{\sigma K}{\lambda K + \sigma} \left( 1 - \frac{\mu}{\lambda K} \right) , \quad
    b_0 = \frac{\mu}{\lambda} \ .  
\end{equation}
The solution $(0,0)$ represents total population extinction. 
At the fixed point $(0,K)$, the predator species goes extinct while the prey species fills the entire system to 
full capacity $K$. 
Finally, the solution $(a_0,b_0)$ with non-zero densities for both species represents predator-prey 
coexistence.
Note that $a_0 > 0$ requires $\mu / \lambda < K$.

Next we consider the (linear) stability of these solutions, which is achieved by linearizing 
(\ref{eq:mean-field}) around the three distinct stationary states. 
Shifting the densities by their stationary solutions $a(t)=a^*+\delta a(t)$, $b(t)=b^*+\delta b(t)$,
inserting this transformation into the original rate equations, and keeping only terms linear in the small 
deviations $(\delta a(t),\delta b(t))$, we obtain the matrix equation $\mathbf{\dot{x}=Jx}$, where 
$\mathbf{x}=\bigl( \delta a(t) \ \delta b(t) \bigr)^T$, the dot represents the time derivative, and the 
Jacobian matrix $\mathbf{J}$ is explicitly given by
\begin{eqnarray}
    \mathbf{J} = \begin{pmatrix} \lambda b^* - \mu & \lambda a^* \\ 
    - \left( \frac{\sigma}{K}+\lambda \right) b^* \  
    & - \lambda a^* + \frac{\sigma}{K} \left( K-a^*-2b^* \right) \end{pmatrix} .
\end{eqnarray}
The dynamical behavior of the system in the vicinity of a fixed point follows from the eigenvalues 
$\epsilon_\pm$ of the Jacobian matrix at each stationary point. 
First let us consider the extinction fixed point $(0,0)$ with associated eigenvalues 
$(\epsilon_-,\epsilon_+) = (- \mu,\sigma)$. 
Both eigenvalues are real, indicating exponential behavior near the fixed point. 
Yet the extinction stationary point is linearly unstable in mean-field theory against prey growth, since 
$\epsilon_+ = \sigma$ is positive. 
While this result is intuitive considering the fact that any small deviation in the prey density leads to 
exponential growth of the prey, we recall that in the original stochastic model, in any finite system 
total extinction represents the only asymptotically stable stationary absorbing state.
Next, the eigenvalues for the predator extinction fixed point $(0,K)$ are 
$(\epsilon_-,\epsilon_+) = (\lambda K - \mu, - \sigma)$, which are also both real.
This stationary state is only stable with respect to small perturbations if $\lambda < \lambda_c = \mu/K$. 
Finally, the two-species coexistence stationary point $(a_0,b_0)$ has associated eigenvalues 
\begin{equation}
   \epsilon_\mp = -\frac{\sigma \mu}{2 \lambda K} \left[ 1 \pm 
   \sqrt{1 - \frac{4 \lambda K}{\sigma} \left( \frac{\lambda K}{\mu} - 1 \right)} \, \right] .
\end{equation} 
For $\lambda_s = (\mu / 2K) \left(1 + \sqrt{1 + \sigma / \mu} \right) > \lambda > \lambda_c$, 
both eigenvalues are real and negative, and hence the stationary point is stable and small perturbations 
exponentially relax back towards it. 
If $\lambda > \lambda_s$, the eigenvalues acquire complex conjugate imaginary components with a 
negative real part, indicating that the stationary point is still stable, but the system exhibits decaying 
oscillations in its vicinity. 
Yet for $\lambda < \lambda_c$ the eigenvalues are both real with $\epsilon_-<0$ and 
$\epsilon_+>0$ assuming opposite signs. 
Consequently, the stationary solution $(a_0,b_0)$ turns into an unstable saddle point.

This analysis demonstrates that the mean-field rate equations (\ref{eq:mean-field}) predict a continuous 
active-to-absorbing state transition at $\lambda = \lambda_c$. 
The absorbing state is the predator extinction phase $(0,K)$ which is stable only for $\lambda < \lambda_c$. 
The active phase is the species coexistence phase $(a_0,b_0)$ which only exists and is then stable for 
$\lambda > \lambda_c$.
This fixed point is a stable node for $\lambda < \lambda_s$ and becomes an attractive focus for 
$\lambda > \lambda_s$. 
The active-to-absorbing phase transition describing predator extinction is also observed in spatially extended
stochastic systems. 
Away from criticality the system's behavior only changes quantitatively relative to the mean-field analysis. 
Near the phase transition the critical exponents governing the model's dynamical scaling laws acquire 
substantial corrections due to fluctuations in dimensions $d \leq d_c = 4$. 
For a more thorough review of the stochastic Lotka--Volterra predator-prey model in a static environment,
we refer to Refs.~\cite{uwe1,uwe2,uwe3}.

\subsection{Periodically switching carrying capacity: numerical integration of the coupled rate equations}
\label{subsection:mean-field_results}

\begin{figure}[t]
    \includegraphics[width=0.95\columnwidth]{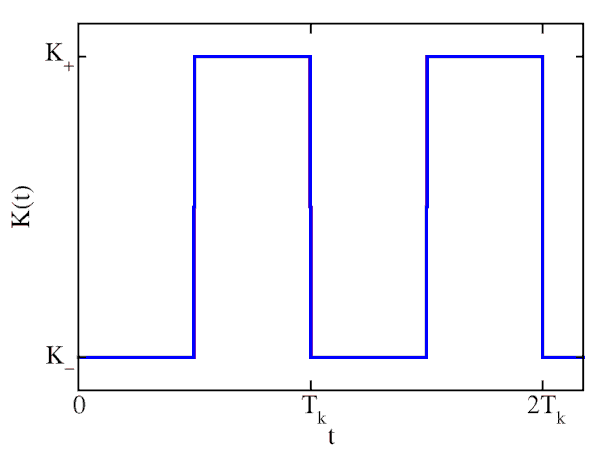}
    \caption{Sketch illustrating the time dependence of the periodically switching carrying capacity $K(t)$: 
    $T_k$ is the full period of the signal; $K_-$ and $K_+$ are its the low and high values.}
\label{fig:K(t)}
\end{figure}
\begin{figure*}[t]
\subfloat[\label{subfig:mean_field_density_a}$T_k=60$]{%
    \includegraphics[width=0.67\columnwidth]{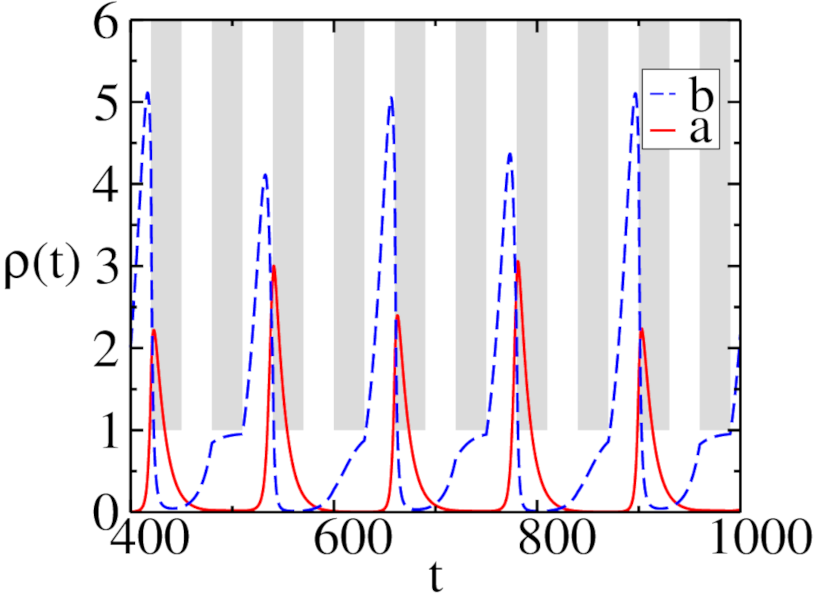} \
}
\subfloat[\label{subfig:mean_field_density_b}$T_k=80$]{%
    \includegraphics[width=0.67\columnwidth]{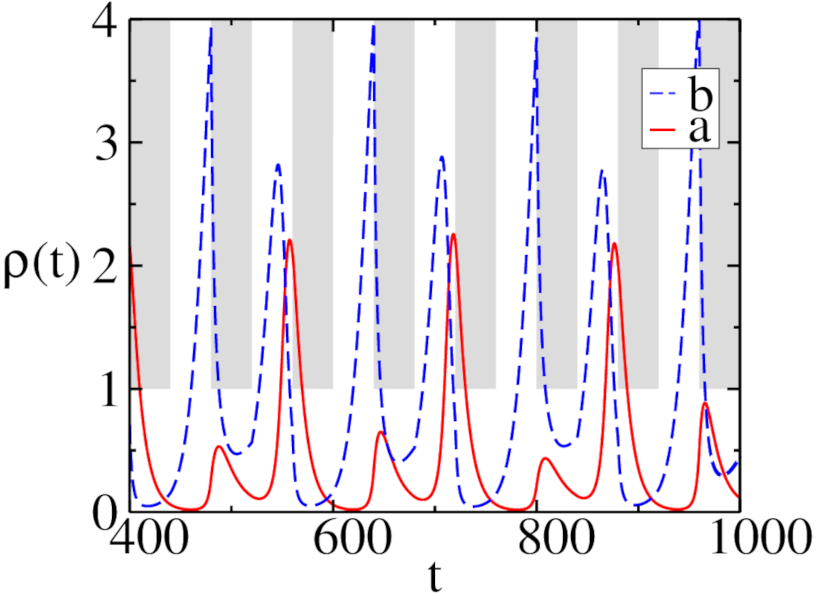} \
}
\subfloat[\label{subfig:mean_field_density_c}$T_k=200$]{%
    \includegraphics[width=0.67\columnwidth]{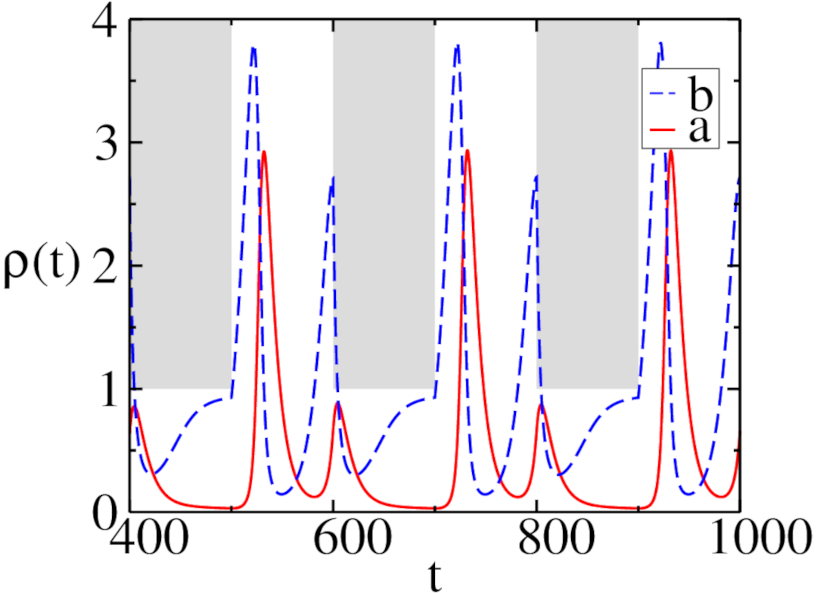}
}
    \caption{Predator (full red) and prey (dashed blue) density time traces obtained by numerical integration 
    of the coupled mean-field rate equations with periodically switching carrying capacity $K(t)$ (the shaded 
    gray areas indicate the excluded densities). 
    The parameters used here are $\sigma=\mu=\lambda=0.1$, and $K_-=1$, $K_+=10$.}
\label{fig:mean_field_density}
\end{figure*}
\begin{figure*}[t]
\subfloat[\label{subfig:mean_field_fft_a}$T_k=60$]{%
    \includegraphics[width=0.67\columnwidth]{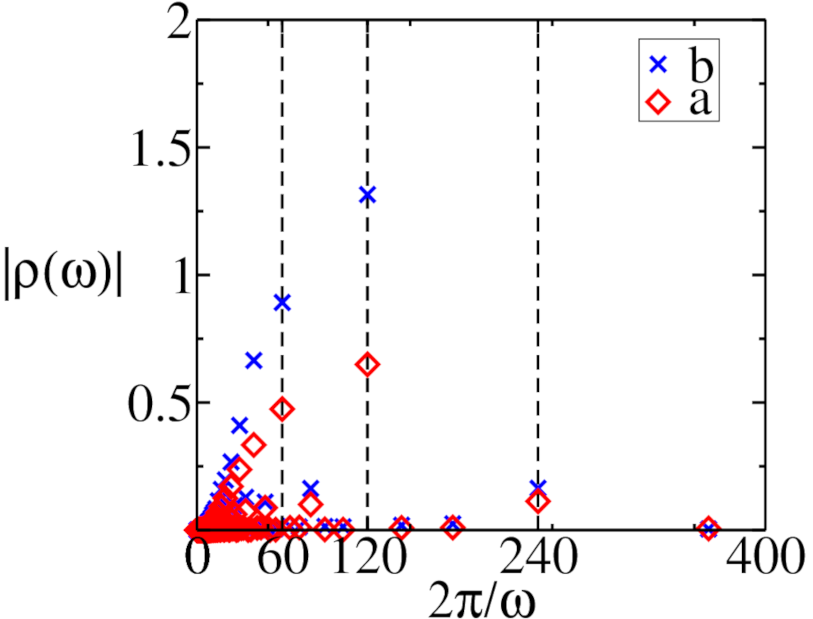} \
}
\subfloat[\label{subfig:mean_field_fft_b}$T_k=80$]{%
    \includegraphics[width=0.67\columnwidth]{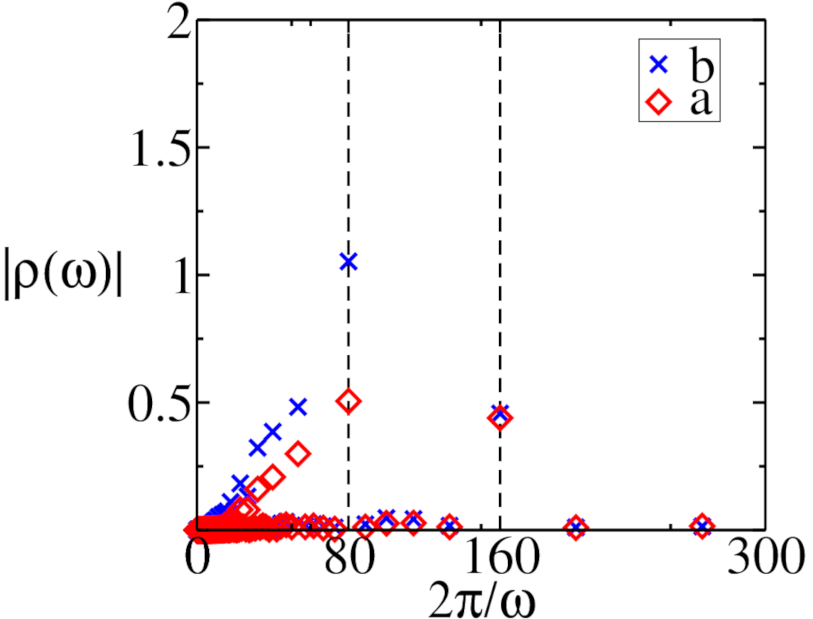} \
}
\subfloat[\label{subfig:mean_field_fft_c}$T_k=200$]{%
    \includegraphics[width=0.67\columnwidth]{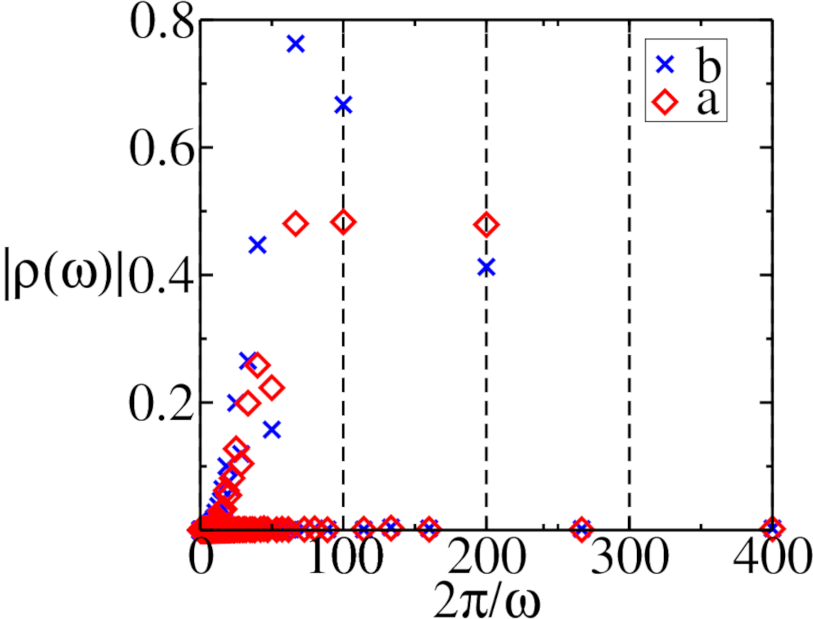}
}
    \caption{Fourier transforms of the predator (red squares) / prey (blue crosses) density time evolution from 
    Fig.~\ref{fig:mean_field_density}, with parameters $\sigma=\mu=\lambda=0.1$, $K_-=1$, $K_+=10$.}
\label{fig:mean_field_fft}
\end{figure*}
In this section, we describe results obtained from numerically integrating the coupled rate equations 
(\ref{eq:mean-field}) subject to a periodically switching carrying capacity.
As depicted in Fig.~\ref{fig:K(t)}, the carrying capacity $K(t)$ is taken to be a rectangular time signal 
ranging between the low and high values $K_-$ and $K_+$, and with full switching period $T_k$ (i.e., 
from $K_\mp$ back to $K_\mp$). 
This functional variation of the carrying capacity does of course not constitute a realistic model for species 
interacting in nature since food resources do not change in a discontinuous manner. 
However, it can be argued that seasonal changes lead to a sudden carrying capacity drop / increase between 
winter and summer, as resource availability may seasonally vary. 
The following results will later be utilized to highlight the differences between the mean-field approximation 
and the stochastic lattice model. 
We remark that a full quantitative comparison between the two models is uninformative due to the fact that in 
the lattice model one prescribes microscopic reaction probabilities, whereas in the mean-field system one 
controls the effective macroscopic reaction rates. 
A thorough quantitative analysis would require fitting the stochastic lattice data to the mean-field results in
order to extract the effective (and usually scale-dependent) macroscopic rates. 
Here, we are not interested in the detailed quantitative differences between the lattice and mean-field models.
Rather we shall focus on the qualitative distinctions between the two models, and will specifically highlight
features predicted by the mean-field equations that are not present in the stochastic lattice 
system.

The mean-field equations were numerically integrated by employing a fourth-order Runge--Kutta scheme with 
(dimensionless) time increment $\Delta t = 0.01$, i.e., $t_0 = 100 \, \Delta t$ sets the basic unit time scale 
relative to which all times and inverse rates will henceforth be measured in this section.
We set the initial conditions to $\rho_a(0) = \rho_b(0) = 0.5$ and $K(0) = K_-$, and have confirmed that our
results do not depend on these chosen initial values. 
Figure~\ref{fig:mean_field_density} displays the resulting predator and prey densities $\rho(t)$ as functions 
of time. 
For both switching periods $T_k=60$ and $T_k=80$, we clearly observe period-doubling effects in the time 
traces.
This is further confirmed by the Fourier transforms of these temporal evolutions shown in 
Fig.~\ref{fig:mean_field_fft}. 
For $T_k=60$, the highest Fourier peak occurs at a period $t=120$ indicating period-doubling. 
However, an additional smaller peak emerges at $t=240$, reflecting that the density repeats after 
four switching periods of the carrying capacity, suggesting even the presence of a period-quadrupling effect. 
Similar period-doubling is visible for $T_k=80$, but no period-quadrupling is discernible. 
Further increase in the carrying capacity period evidently eliminates period-doubling phenomena as shown in 
Fig.~\ref{subfig:mean_field_fft_c}. 
We detect the highest peak in the density Fourier transforms for $T_k=200$ at $t = T_k / 3$, a harmonic of 
the driving period. 
This feature is in fact also observed in the lattice model, in contrast to the period-doubling at smaller 
periods $T_k$, for which we shall find that the internal reaction noise in the stochastic model washes away 
these intriguing non-linear effects.

\subsection{Quantitative analysis: slow-switching regime}
\label{subsection:quantative_analysis_slow}

The stationary mean-field population densities in the coexistence phase are given in 
Eq.~(\ref{eq:stationary}).
For an environment where $K$ periodically switches between two constant values $K_-$ and $K_+$, the 
long-time behavior of the system depends on these stationary densities. 
If the period of the oscillating environment is sufficiently long such that the system reaches the stationary 
state for either $K$ value, then the densities can effectively be described as oscillating between two constant 
values with the same period $T_k$ as the carrying capacity. 
In that case, the averages of the predator and prey densities over one period can simply be approximated by 
the arithmetic means $(\Tilde{a},\Tilde{b})$ of the two stationary values $(a_-,b_-)$ and $(a_+,b_+)$
pertaining to $K=K_-$ and $K=K_+$, respectively. 
Thus we obtain
\begin{subequations}
    \begin{eqnarray}
    \Tilde{a} &=& \frac{a_-+a_+}{2} = \frac{\sigma}{2 \lambda} \left( \frac{\lambda K_- - \mu}
    {\lambda K_- + \sigma} + \frac{\lambda K_+ - \mu}{\lambda K_+ + \sigma} \right) 
    \label{subeq:averagepredator} \nonumber \\
    &=&  \frac{\sigma}{\lambda} \ \frac{2 \lambda^2 K_ - K_+ + \lambda (\sigma - \mu) (K_- + K_+) 
   - 2 \mu \sigma}{2 (\lambda K_- + \sigma) (\lambda K_+ + \sigma)} \, , \ \\
    \Tilde{b} &=& \frac{b_-+b_+}{2} = \frac{\mu}{\lambda} \ .
    \end{eqnarray}
\end{subequations}
We rewrite the mean predator density in terms of an equivalent time-averaged effective carrying capacity 
$K^*$ defined through
\begin{equation}
    \Tilde{a} = \frac{\sigma}{\lambda} \ \frac{\lambda K^*-\mu}{\lambda K^*+\sigma} \ .
    \label{eq:a_tilde}
\end{equation}
Comparison with the explicit result (\ref{subeq:averagepredator}) yields
\begin{equation}
    K^* = \frac{2 K_- K_+ + (K_- + K_+) \, \sigma / \lambda}{K_- + K_+ + 2 \sigma / \lambda} \ ,
    \label{eq:equivalent_K}
\end{equation}
which reduces to the rate-independent harmonic average ${\bar K} = 2 K_ - K_+ / (K_- + K_+)$ for large 
$K_-, K_+ \gg 1$.
Hence, in the slow-switching regime, the system can be described as oscillating around the average 
population densities corresponding to the constant rate-dependent effective carrying capacity $K^*$.

\begin{figure}[t]
\subfloat[\label{subfig:mean_field_K_eq_a}predator density]{%
    \includegraphics[width=0.5\columnwidth]{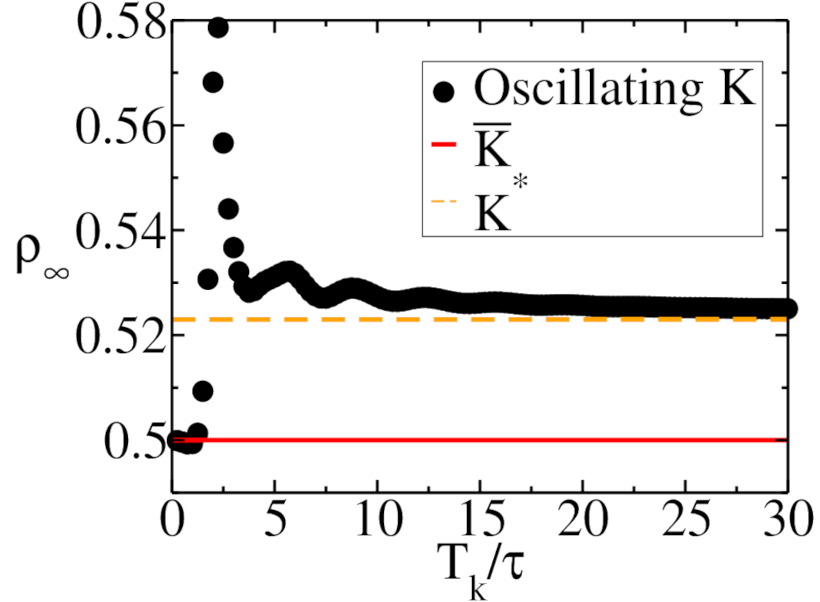}%
}\hfill
\subfloat[\label{subfig:mean_field_K_eq_b}prey density]{%
    \includegraphics[width=0.5\columnwidth]{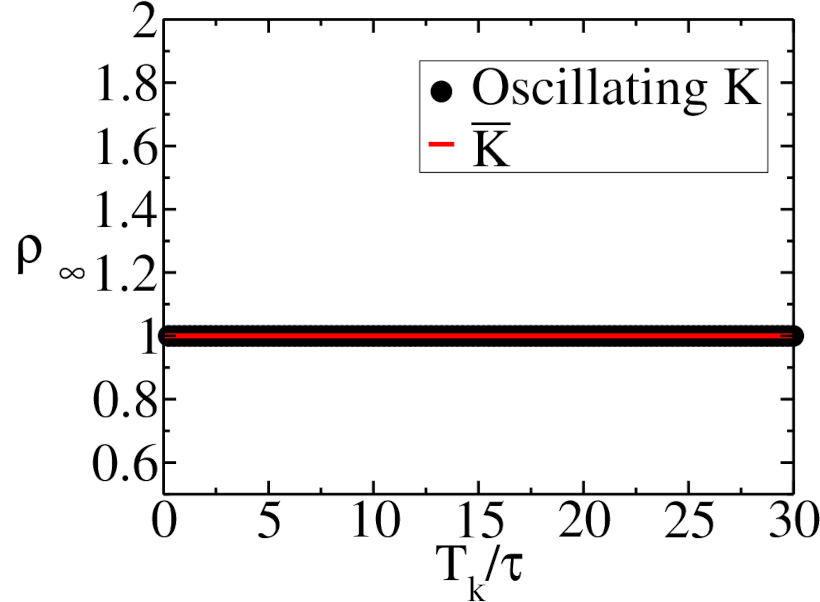}%
}
    \caption{Long-time predator and prey densities $\rho_\infty$ averaged over $10$ periods of the switching 
    carrying capacity, vs. $T_k$ in units of $\tau$, where $\tau$ represents the characteristic intrinsic oscillation
    period for a Lotka--Volterra model with fixed carrying capacity $\bar{K}$. 
    The parameters used for the oscillating environment are $\sigma=\mu=\lambda=0.1$, and $K_-=2$, 
    $K_+=6$, which yields the harmonic average $\bar{K} = 3$ (red) and $K^*=3.2$ (dashed orange). 
    The corresponding stationary densities follow from Eq.~(\ref{eq:stationary}).}
\label{fig:mean_field_K_eq}
\end{figure}
\begin{figure}[t]
    \includegraphics[width=0.95\columnwidth]{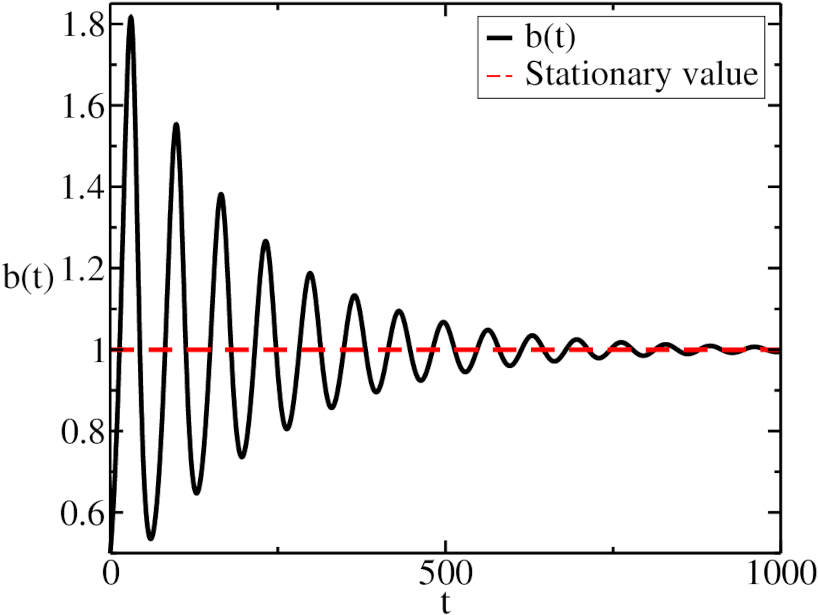}
    \caption{Numerical integration for the prey density $b(t)$ for a static environment (full black) for 
    $\sigma = \mu = \lambda = 0.1$ and $K = 10$, compared with the stationary value $b_0 = 1$ 
    (dashed red). 
    The average of the oscillating black curve over time is ${\bar b} = 1.00246$.}
    \label{fig:prey_average_density}
\end{figure}

Through numerical integration of the mean-field rate equations, we tested the harmonic average 
hypothesis for different switching periods, and confirmed Eq.~(\ref{eq:equivalent_K}) in the slow-switching
regime. 
We note that this comparison is facilitated for the mean-field model compared to the stochastic lattice system
because we have exact formulas available for the stationary density values, and $K$ is not required to be an
integer. 
Figure~\ref{fig:mean_field_K_eq} shows the comparison of the numerically obtained population densities
with periodically varying $K(t)$ with the corresponding stationary values obtained with a simple harmonic 
average of the carrying capacities and the rate-dependent effective carrying capacity (\ref{eq:equivalent_K}).
Interestingly, computing the stationary prey density from the straightforward harmonic carrying capacity 
average ${\bar K}$ yields accurate results for a large range of switchting periods, as is apparent in 
Fig.~\ref{subfig:mean_field_K_eq_b}.
This is due to the fact that for a static carrying capacity, the prey density oscillates about its stationary value,
and the fluctuations about it almost precisely average out, as verified in Fig.~\ref{fig:prey_average_density}. 
Since within the mean-field framework, $b^*$ and hence $\Tilde b$ do not depend on $K$, any equivalent 
carrying capacity would work for the prey population.
The predator density also follows the harmonically averaged carrying capacity for small periods, see below;
and is indeed aptly captured by the rate-dependent equivalent carrying capacity (\ref{eq:equivalent_K}) for
large switching periods.
For intermediate periods $T_k$, we observe a non-monotonic crossover regime with a large resonance-like
spike, see Fig.~\ref{subfig:mean_field_K_eq_a}. 
We verified that these findings do not depend on the initial conditions of the system.

\subsection{Quantitative analysis: fast-switching regime}
\label{subsection:quantative_analysis_fast}

The coupled mean-field rate equations (\ref{eq:mean-field}) suggest that in the fast-switching regime, both
species' densities oscillate about values that are equal to the stationary population densities for an equivalent
carrying capacity ${\bar K}$ that is just the harmonic average of $K_-$ and $K_+$. 
This follows from the fact that the prey density rate equation (\ref{subeq:pred_MF}) depends explicitly on 
$1 / K$. 
Based on these observations, we construct an ansatz for the long-time behavior of both species' densities as 
follows.

We first shift time according to $t \to t - NT_k$, where $N$ is a large integer such that at $t = N T_k$ the 
system has reached its quasi-stationary state. 
Hence this time axis shift defines $t=0$ to be the start of an environmental cycle in the long-time regime. 
If the system is thus initialized at the onset of the low carrying capacity state, i.e., at $t = 0$ it just switched 
from $K_+$ to $K_-$, then at $t = T_k / 2$ it will flip back from $K_-$ to $K_+$, and that cycle repeats at 
$t =T_k$. 
We now derive an approximate solution that describes the densities in one cycle $t \in [0,T_k]$. 
Henceforth we shall refer to the region $t \in [0, T_k / 2]$ as $\mathcal{T_-}$, and the time interval
$t \in [T_k / 2,T_k]$ as $\mathcal{T_+}$.

\begin{figure}
    \includegraphics[width=0.95\columnwidth]{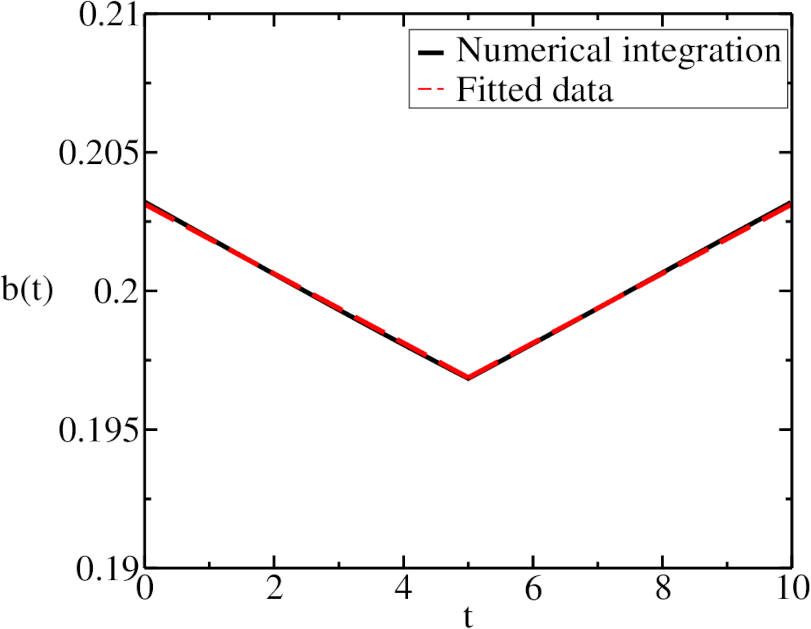}
    \caption{Numerical integration of the mean-field equations (black) for the parameters 
    $\sigma = \mu = 0.1$, $\lambda = 0.5$, $T_k = 10$, and $K_- = 2$, $K_+ = 6$. 
    The dashed red graph represents a linear fit applied to the numerical data, resulting in $\alpha = 0.00125$.}
    \label{fig:prey_ansatz}
\end{figure}

Since the prey density exhibits a discontinuity in its first time derivative at $t = T_k / 2$, it can be described 
by a piece-wise function. 
In the fast-switching regime, we may apply a short-time Taylor expansion for the population dynamics, and
retain only the linear term. 
The absolute values of the prey density slope in the intervals $\mathcal{T_-}$ and $\mathcal{T_+}$ must 
be the same, due to the fact that the prey density is periodic, $b(T_k) = b(0)$, and continuous at the jumps
in between these two regions. 
In $\mathcal{T_-}$ the system is in the low carrying capacity state, therefore the prey density is a decreasing 
function of time, and its slope should be negative. 
For $t \in \mathcal{T_+}$, the prey density has a positive slope, since now the system is in the high carrying 
capacity state. 
These considerations motivate the following simple ansatz for the prey density,
\begin{eqnarray}
\label{eq:prey_ansatz}
   b(t) = \begin{cases}
            b_1-\alpha t & \quad t \in \mathcal{T_-} \ , \\
            b_2+\alpha t & \quad t \in \mathcal{T_+} \ , \end{cases}
\end{eqnarray}
which is numerically verified in Fig.~\ref{fig:prey_ansatz}.

The prey density is continuous at the boundary $t = T_k / 2$, whence $b_2 = b_1 - \alpha T$. 
Moreover, in the fast-switching regime, the density variations of both species over one period of the carrying
capacity should be assumed to be small relative to their average values. 
Consequently, $1 / K(t)$ is the only significant term when averaging Eq.~(\ref{subeq:prey_MF}). 
Its average leads to an equivalent carrying capacity that is equal to the harmonic average ${\bar K}$. 
Therefore, the system reaches a quasi-stationary state, where both densities oscillate around their stationary 
values for an equivalent carrying capacity ${\bar K}$. 
The temporal average of Eq.~(\ref{eq:prey_ansatz}) needs to be $b_0$. 
Imposing this condition, we obtain
\begin{eqnarray}
\label{eq:prey_ansatz2}
   b(t) = \begin{cases}
            b_0 - \alpha \left( t - \frac{T_k}{4} \right) & \quad t \in \mathcal{T_-}  \ , \\
            b_0 + \alpha \left( t - \frac{3T_k}{4} \right) & \quad t \in \mathcal{T_+} \ ; \end{cases}
\end{eqnarray}
the slope constant $\alpha$ shall be determined later.

The rate equation for the predator density (\ref{subeq:pred_MF}) may now be cast into a more suggestive 
form,
\begin{equation}
    \dot{a}(t) = \lambda \, a(t) \left[ b(t) - b_0 \right] ,
    \label{eq:modified_pred_MF}
\end{equation}
which indicates that the extrema of $a(t)$ occur at times when $b(t) = b_0$. 
Using the ansatz~(\ref{eq:prey_ansatz2}), this happens at $t = T_k / 4$ and $t =  3T_k / 4$. 
Equation~(\ref{eq:modified_pred_MF}) can then be integrated to solve for the predator density,
\begin{eqnarray}
    a(t) &\sim& A \, e^{\lambda \left( \int^t b(t') \, dt' - b_0 t \right)} \nonumber \\ 
    &=& \begin{cases}
            A \, e^{- \frac{\lambda \alpha}{2} \left( t^2 - \frac{T_k}{2} t \right)} 
    		& \quad t \in \mathcal{T_-} \ , \\
            A' \, e^{\frac{\lambda \alpha}{2} \left(t^2 - \frac{3T_k}{2} t \right)} 
		& \quad t \in \mathcal{T_+} \ , \end{cases} \nonumber
\end{eqnarray}
where $A$ and $A'$ are integration constants. 
Since the predator density is required to be continuous at $t = T_k / 2$, one arrives at the relation 
$A' = A \, e^{\lambda\alpha T_k^2 / 4}$, which yields the approximate predator density solution
\begin{eqnarray}
    a(t) = \begin{cases}
            A \, e^{- \frac{\lambda\alpha}{2} \left( t^2 - \frac{T_k}{2} t \right)} 
		& \quad t \in \mathcal{T_-} \ , \\
            A \, e^{\frac{\lambda\alpha}{2} \bigl( t^2 - \frac{3T_k}{2} t + \frac{T_k^2}{2} \bigr)} 
		& \quad t \in \mathcal{T_+} \ . \end{cases}
	\label{eq:aans}
\end{eqnarray}
The average of the predator density over one cycle of environmental switching then becomes
\begin{eqnarray}
    \frac{1}{T_k} \int_0^{T_k} \!\! a(t) \, dt = 2\sqrt{2 \pi} A \, e^{T_k^2 \alpha \lambda / 32} \ 
    \frac{\erf\!\left( \frac{T_k \sqrt{\alpha\lambda}}{4 \sqrt{2}} \right)}{T_k \sqrt{\alpha\lambda}} \ . \
    \label{eq:average_pred}
\end{eqnarray}

Under the assumption of fast environmental switching, $T_k$ should be the smallest time scale in the 
system, and the explicit form of Eq.~(\ref{eq:average_pred}) suggests that the fast-switching regime is 
quantitatively delineated by $T_k \sqrt{\alpha\lambda} \ll 1$. 
The still undetermined parameter is the (initial) slope of the prey density $\alpha = |\dot b|_0$.
To zeroth order in $\alpha$, either immediately from Eq.~(\ref{eq:aans}) or by expanding 
Eq.~(\ref{eq:average_pred}) in $T_k \sqrt{\alpha\lambda}$, gives the simple result
\begin{eqnarray}
    \frac{1}{T_k}\int_0^{T_k} a(t) \, dt = A + O(T_k \sqrt{\alpha\lambda}) \ .
\end{eqnarray}
Since this average must equal the stationary value of predator density for a harmonically averaged carrying 
capacity, we may fix the integration constant
\begin{eqnarray}
    A \approx \frac{\sigma}{\lambda} \, \frac{\lambda {\bar K} - \mu}{\lambda {\bar K} + \sigma} =
    \frac{\sigma}{\lambda} \, \frac{2 \lambda K_+K_- - \mu \left( K_++K_- \right)}
   {2 \lambda K_+ K_- + \sigma \left( K_+ + K_- \right)} \, , \
   \label{eq:referee1}
\end{eqnarray}
to leading order in an expansion in powers of $T_k \sqrt{\alpha\lambda}$.

The left-hand side of Eq.~(\ref{subeq:prey_MF}) equals the constant slope of the prey density under the 
fast-switching approximation. 
Since $t < T_k $, we also have $t \sqrt{\alpha\lambda} \ll 1$, and with 
$a(t) = A +O(T_k \sqrt{\alpha\lambda})$ one has $b(t) = b_0 + O(T_k \alpha)$.
Upon inserting these asymptotic values into (\ref{subeq:prey_MF}) for $t \in \mathcal{T_-}$, we arrive at
\begin{equation}
    -\alpha\approx\sigma b_0\left(1-\frac{A+b_0}{K_-}\right)-\lambda A b_0 \ ,
\end{equation}
and thus inserting Eq.~(\ref{eq:referee1}) we obtain
\begin{eqnarray}
    \alpha \approx \frac{\mu\sigma}{\lambda} \, 
    \frac{(\mu + \sigma) (K_+ - K_-)}{2\lambda K_+ K_- + \sigma (K_++K_-)} \ .
\end{eqnarray}
These approximations fully characterize the long-time quasi-stationary state in the fast-switching regime.

\begin{figure}[t]
    \includegraphics[width=0.95\columnwidth]{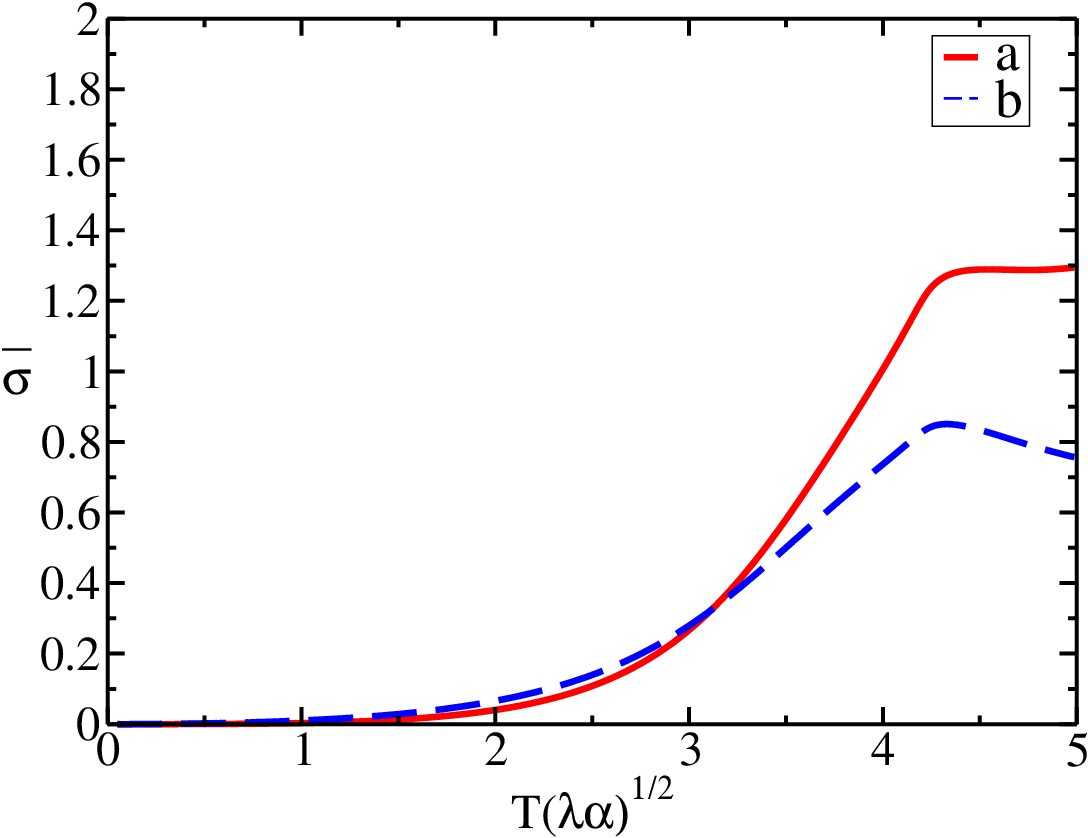}
    \caption{Relative root mean-square error of the approximate solution as function of
    $T_k \sqrt{\lambda\alpha}$ for the predator (full red) and prey (dashed blue) populations.}
    \label{fig:error_plot}
\end{figure}
In order to test this approximate solution, we computed the root mean-square error between our ansatz and 
the result of numerically integrating the mean-field equations. 
This error was then divided by the actual density average as obtained from numerical integration to obtain a 
dimensionless error measure $\bar{\sigma}$. 
In Fig.~\ref{fig:error_plot}, this relative error $\bar{\sigma}$ is plotted against the dimensionless carrying 
capacity period $T_k \sqrt{\lambda\alpha}$. 
As expected, our approximation yields small relative errors for $T_k \sqrt{\lambda\alpha} \ll 1$. 
Interestingly, the asymptotic expansion seems to work even up to values $T_k \sqrt{\lambda\alpha} = 2$ 
with relative errors less than $10 \%$.

\section{Stochastic lattice model}
\label{section:lattice_model}

\subsection{Stochastic Monte Carlo simulation algorithm}
\label{subsection:methods}

In this section, we employ a lattice model to numerically simulate the stochastic Lotka--Volterra 
predator-prey system (\ref{eq:reactions}), which allows us to investigate spatial structures and 
reaction-induced spatio-temporal correlations. 
Utilizing a stochastic lattice model allows us to investigate resonance effects on correlations, 
and their relation to the intrinsic spatio-temporal patterns of our system. 
Direct comparison with the mean-field rate equation approximation delineates the latter's validity range, thus
providing information when stochastic fluctuations and correlation effects may be ignored without losing
pertinent qualitative features.
The stochasticity of the system is implemented through an individual-based Monte Carlo algorithm. 
We implement the model on a two-dimensional square lattice with periodic boundary conditions (i.e., a 
toroidal simulation domain), where each lattice site holds information about the number of individuals of 
each species at that location. 
The initial configuration of the system is set up as a disordered state where each individual is placed at a 
randomly selected lattice site. 
We employ the following notation:
$n_a(x,y;t)$: number of predator individuals at site $(x,y)$ and at time $t$; $n_b(x,y;t)$: number of prey 
individuals at site $(x,y)$ and at time $t$; $N_a(t)$: total number of predator individuals across the entire 
lattice at time $t$; $N_b(t)$: total number of prey individuals across the entire lattice at time $t$; 
$\bar{n}_i(t)$: $N_i(t) / L^2$, where $L$ is the linear lattice size, denotes the average species $i \in (a,b)$ 
density.
Time is simulated via Monte Carlo steps (MCS), such that at each Monte Carlo time step
\begin{enumerate}
    \item a random location on the lattice $(x,y)$ is picked;
    \item a random neighboring site is selected from the von-Neumann neighborhood (four nearest 
    neighbors) $(x_{\rm new},y_{\rm new})$;
    \item if $(x,y)$ contains a predator individual, we attempt $n_b(x_{\rm new},y_{\rm new};t)$ 
    predation reactions as follows:
    \begin{itemize}
        \item generate a uniformly distributed random number $r$;
        \item if $r<\lambda$, decrease the number of prey at $(x_{\rm new},y_{\rm new})$ by $1$ and 
        increase the number of predators at $(x_{\rm new},y_{\rm new})$ by $1$;
    \end{itemize}
    \item next attempt a death reaction for the predator as described below:
    \begin{itemize}
        \item generate a uniformly distributed random number $r$.
        \item if $r < \mu$, decrease the number of predators at $(x,y)$ by $1$;
    \end{itemize}
    \item if $(x,y)$ contains a prey individual, attempt a reproduction reaction as follows:
    \begin{itemize}
        \item generate a uniformly distributed random number $r$;
        \item if $r < \sigma$ and $n_a(x_{\rm new},y_{\rm new};t) \\ 
        +n_b(x_{\rm new},y_{\rm new};t)<K$,
        increase the number of prey at site $(x_{\rm new},y_{\rm new})$ by $1$;
    \end{itemize}
    \item if $(x,y)$ is empty ($n_a(x,y;t)+n_b(x,y;t)=0$), return to step 1;
    \item the above steps are repeated $N_a(t)+N_b(t)$ times.
\end{enumerate}
\begin{figure*}
\subfloat[\label{subfig:stochastic_density_movie_a}$t=0$]{%
  \includegraphics[width=0.5\columnwidth]{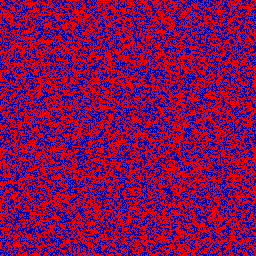}%
}\hfill
\subfloat[\label{subfig:stochastic_density_movie_b}$t=46$]{%
  \includegraphics[width=0.5\columnwidth]{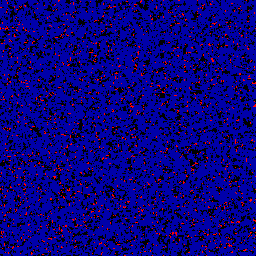}%
}\hfill
\subfloat[\label{subfig:stochastic_density_movie_c}$t=70$]{%
  \includegraphics[width=0.5\columnwidth]{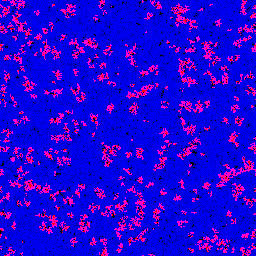}%
}\hfill
\subfloat[\label{subfig:stochastic_density_movie_d}$t=88$]{%
  \includegraphics[width=0.5\columnwidth]{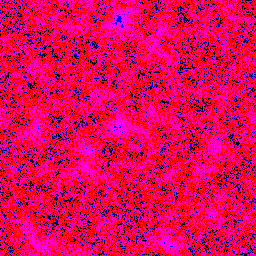}%
}
\caption{Snapshots of a single run for a system with parameters $L = 256$, 
$\sigma = \mu = \lambda = 0.1$, $K_- = 1$, $K_+ = 10$, and $T_k = 100$; time $t$ is measured in 
units of Monte Carlo steps. 
The red and blue colored pixels indicate the presence of predators and prey, respectively, with the 
brightness representing the local density, the pink colored pixels pertain to sites with both predator and prey
present, and the black pixels represent empty sites. 
The system is initialized with $K(t=0) = K_- = 1$. 
The full movie can be viewed at the link provided in Ref.~\cite{note:movies}.}
\label{fig:stochastic_density_movie}
\end{figure*}
This implementation ensures that at each Monte Carlo time step, on average, all individuals in the lattice 
attempt a reaction. 
We utilize random updates (i.e., picking new lattice sites at random) rather than systematic sequential updates 
(going over each lattice site in a specific sequence) in order to avoid introducing any bias in how reactions 
occur in the system.

A choice now has to be made in how to precisely manage the population after switching from the high to the 
low carrying capacity, because there will likely be an excess number of individuals at some lattice sites. 
We have considered two implementations to deal with this issue:
In the first variant, we randomly removed any excess individuals to immediately reach the allowed low 
carrying capacity value $K_-$. 
While this implementation leads to interesting period-doubling behavior, we deemed it to be unrealistic. 
In the second implementation, we left the excess particles on site, but restricted further prey reproduction at 
lattice locations with more individuals than permitted. 
Therefore, we allow the system to intrinsically relax to a configuration without excess individuals, since 
eventually any superfluous predators would be forced to perish, and any excess prey would be devoured by 
predators. 
This intrinsic relaxation introduces a time scale set by the internal response time of the system, which is in 
turn determined by the reaction rates.

The stochastic lattice system was simulated over multiple runs, thus averaging both over ensembles of
different initial conditions and distinct temporal histories; $\langle \ldots \rangle$ denotes the resulting 
(double) ensemble averages. 
We measured the average spatial species densities $\rho_i(t) = \langle \bar{n}_i(t) \rangle$, and computed 
the (connected) auto-correlation functions at fixed positions, 
$C_{ij}(t,t_0) = \langle \bar{n}_i(t)\bar{n}_j(t_0) \rangle 
- \langle \bar{n}_i(t) \rangle \langle \bar{n}_j(t_0) \rangle$. 
The static correlations as functions of spatial distance $|x - x_0|$ were extracted using the definition
$C_{ij}(x,x_0;y_0,t_0) = \langle n_i(x,y_0,t_0) n_j(x_0,y_0,t_0) \rangle - \langle n_i(x,y_0,t_0) \rangle 
\langle n_j(x_0,y_0,t_0) \rangle$. 
In the long-time regime, the system should be isotropic at length scales large compared with the lattice 
constant, so that static correlations along the $x$ or $y$ directions will become identical. 
We also assume that the system is homogeneous at those scales, and hence that the auto-correlations 
should be independent of the reference positions $(x_0,y_0)$. 
Consequently we determine the auto-correlations using the densities averaged over lattice sites, which
improves our statistics. 
Both these assumption were confirmed via explicit simulations. 
Furthermore, we evaluated the static correlations at a specific, sufficiently late fixed time step $t_0$, but
again checked that all correlations are invariant under discrete time translation $t_0 \to t_0 +T_k$ with the
environmental switching period $T_k$.

The various parameters of the system are the three reaction rates $(\sigma,\mu,\lambda)$, the low and high 
carrying capacity values $(K_-,K_+)$, and the period of the oscillating environment $T_k$. 
However, we can eliminate one of these parameters by rescaling the units of time. 
In our Monte Carlo simulations, which are not intended to match any specific experimental
or observational data, we chose to always fix $\sigma = \mu$, because these parameters represent the 
rates for the linear prey reproduction and predator death reactions, and we are predominantly interested in the 
behavior of the system as the non-linear coupling $\lambda$ is varied. 
Thus our varying control parameters consist of the set $(\lambda, K_-, K_+, T_k)$. 
For fixed $\sigma$ and $\mu$, the critical predation rate $\lambda_c$ only depends on the carrying capacity. 
Therefore, for the remainder of this paper we shall implicitly assume a fixed value for $\sigma = \mu$ and 
indicate the critical threshold as $\lambda_c(K)$.

\subsection{Population densities}
\label{subsection:density}

Snapshots of a single simulation run of a representative system at different times are depicted in 
Fig.~\ref{fig:stochastic_density_movie}. 
According to our setup, the system is initially in a random configuration, so the predators consume the prey 
available in their neighborhood. 
At $t=46$, the predators have devoured most of their prey, and their number decays over time: 
The system is in the predator extinction phase for $K=1$. 
Therefore, without the external periodic environmental variation, the predators would eventually go extinct. 
However, we see that at $t = 70$, after the carrying capacity has jumped at $t = T_k/2 =50$ from $K = 1$ 
to $K = 10$, the prey are permitted to reproduce more abundantly. 
The prey population increase induces spreading waves of predators, in turn causing an enhancement of the
predator density, until by $t=88$ the latter almost fill the entire lattice.
When the carrying capacity drops back to $K = 1$ at $T_k = 100$, the predator density starts decaying again
over time towards the point of extinction until the carrying capacity is once more reset and the whole process 
(stochastically) repeats.

\begin{figure}
\subfloat[\label{subfig:stochastic_density_a}$\lambda=0.1$]{%
  \includegraphics[width=0.5\columnwidth]{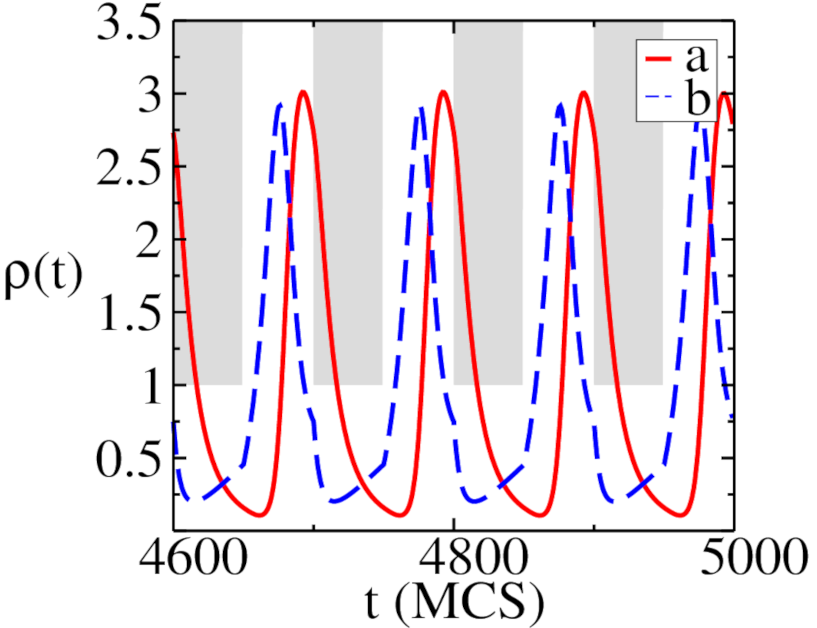}%
}\hfill
\subfloat[\label{subfig:stochastic_density_b}$\lambda=0.275$]{%
  \includegraphics[width=0.5\columnwidth]{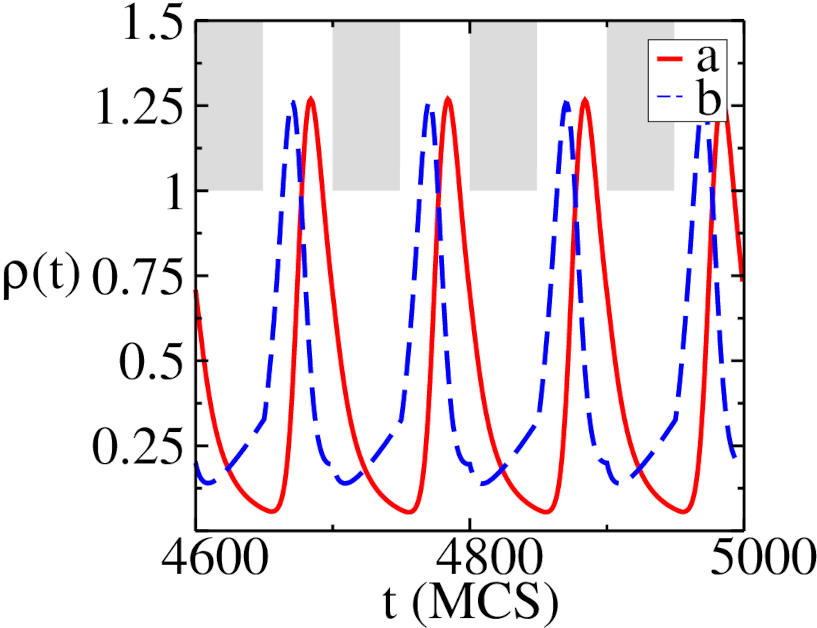}%
}
\caption{Predator (full red) and prey (dashed blue) population densities averaged over $50$ realizations for
a system with $L = 256$, $\sigma = \mu = 0.1$, $K_- = 1$, $K_+ = 10$, and $T_k = 100$; the shaded 
gray areas are excluded by the switching carrying capacity $K(t)$. 
The critical predation rate values associated with fixed carrying capacities $K_-$, $K_+$ are 
$\lambda_c(K=1) = 0.26(5)$ and $\lambda_c(K=10) = 0.01(0)$.}
\label{fig:stochastic_density}
\end{figure}
\begin{figure}[b]
    \includegraphics[width=0.95\columnwidth]{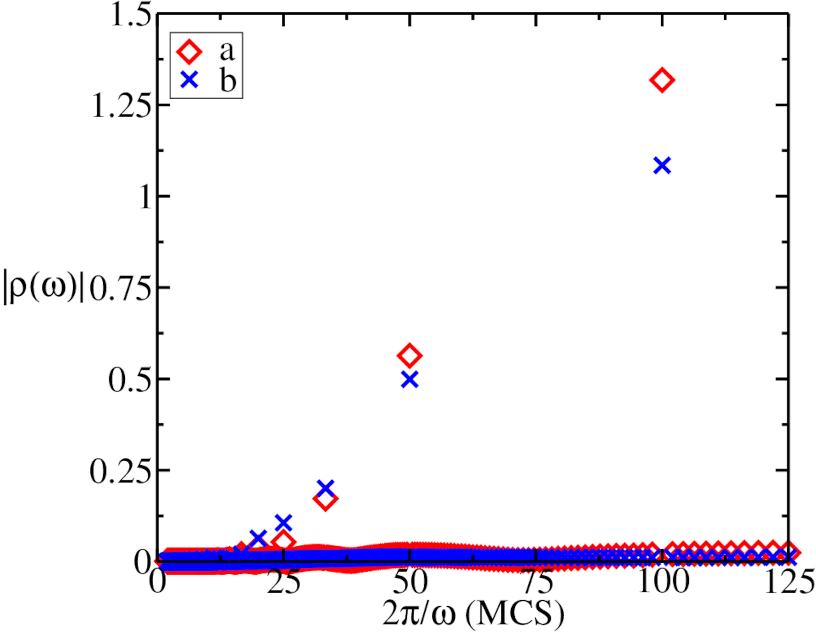}
    \caption{Population density Fourier transforms as functions of the period $2 \pi / \omega$ (predators:
    red squares; prey: blue crosses). 
    The first $2000$ Monte Carlo time steps were discarded before computing the Fourier transform in order to 
    eliminate the initial behavior. 
    The parameters used here are $L = 256$, $\sigma = \mu = \lambda = 0.1$, $K_- = 1$, $K_+ = 10$, and 
    $T_k = 100$.}
    \label{fig:stochastic_fft}
\end{figure}
Figure~\ref{fig:stochastic_density} shows the long-time behavior of the density for two different values of 
the predation rate $\lambda$. 
For $\lambda=0.1$ (a), the system oscillates between the predator extinction phase, approached when 
$K = 1$, and the two-species coexistence phase, when $K = 10$. 
In contrast, for $\lambda = 0.275$ (b) the system resides in the species coexistence phase at both $K$ values. 
Both population time traces show stable oscillations with the switching period $T_k$, as expected for a 
dynamical system driven by a periodic external force. 
This is further confirmed by the Fourier transform plots displayed in Fig.~\ref{fig:stochastic_fft}. 
The prey density becomes non-smooth at points where the carrying capacity switches from $K_+$ to $K_-$ or 
vice versa, while the predator density remains smooth at those points. 
This is indicative of the fact that only the prey density explicitly depends on the carrying capacity, while the 
predator density depends on $K$ through its coupling to the prey species. 
In Fig.~\ref{subfig:stochastic_density_movie_a}, we see that even though the system is in the predator 
extinction phase when $K=1$, the $A$ species are still able to maintain a non-zero population density through 
the periodic environmental variation. 
Indeed, we observe that the key difference between the runs for $\lambda = 0.1$ and $\lambda = 0.275$ 
resides in the amplitude of the oscillations, which drops significantly when the predation rate increases. 
This is a general feature of the static Lotka--Volterra model. 
However, the amplitude of the oscillation in Fig.~\ref{subfig:stochastic_density_movie_a} is even higher than 
would be attained in a static system with fixed $K = 10$:
Driving the system away from reaching the absorbing state causes the densities to overshoot their stationary 
state values for $K=10$.
While a static system would go extinct for low values of the predation rate, the periodic temporal variation of 
the carrying capacity allows both species to coexist in this situation. 

\begin{figure}[t]
\subfloat[\label{subfig:better_coexistence_a}predators]{%
  \includegraphics[width=0.5\columnwidth]{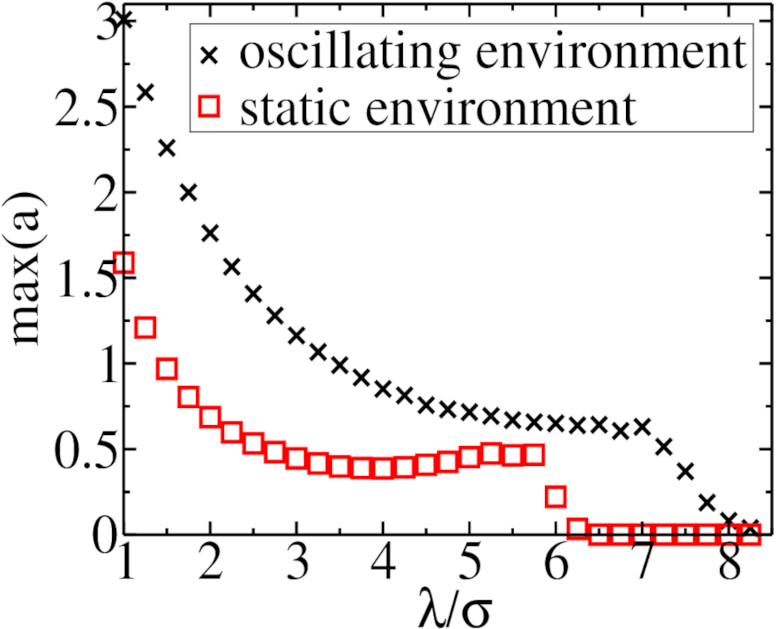}%
}\hfill
\subfloat[\label{subfig:better_coexistence_b}prey]{%
  \includegraphics[width=0.5\columnwidth]{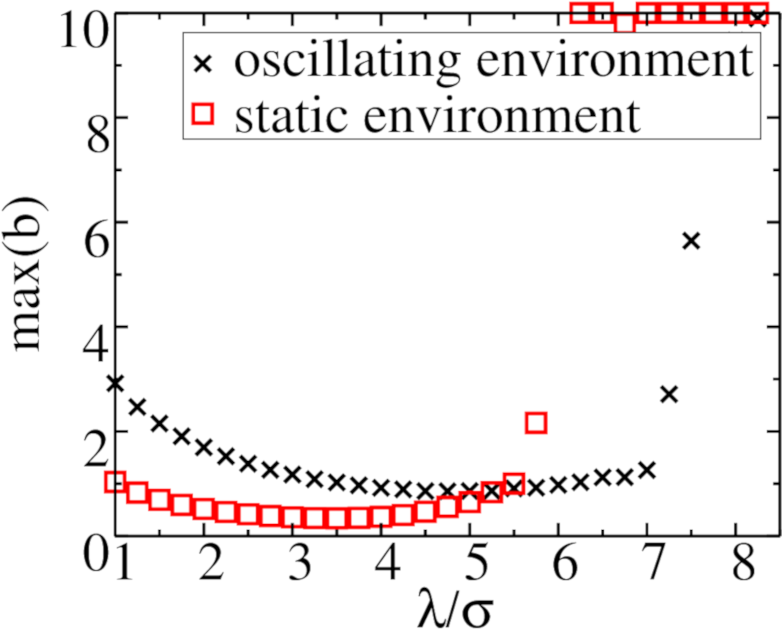}%
}
\caption{Maximum population densities achieved in the late time interval $t \in [2000,5000]$, plotted 
against the dimensionless rate ratio $\lambda / \sigma$, where $L = 256$, $\sigma = \mu = 0.1$, 
$K_+ = 10$, $K_- = 1$, and $T_k = 100$ for the oscillating environment (black crosses), and for the same 
parameters with fixed $K = 10$ for the static case (red squares).}
\label{fig:better_coexistence}
\end{figure}
In fact, the population oscillations become most prominent if the carrying capacity effectively switches the 
system between the predators' absorbing and active phases. 
To demonstrate that this is a generic feature of our model, we plot the maximum density values reached in the
simulations in the long-time limit in Fig.~\ref{fig:better_coexistence}. 
For $\lambda > 6 \sigma$, we observe predator extinction; as was noted in Ref.~\cite{physics1}, the system 
may, depending on the initial conditions, evolve into one of the two absorbing states for large predation rates. 
We interpret this extinction transition to be caused by stochastic fluctuations in our finite simulation system:
As the predation rate becomes large, stochastic fluctuations are increasingly likely to drive the simulation 
towards the absorbing predator extinction state.
For smaller predation rates, the asymptotic predator density decreases with growing $\lambda$.
In the two-species coexistence region, the simulation results for the systems with periodically varying 
environment exhibit markedly larger oscillation amplitudes for both predatator and prey populations. 
This enhancement of the maximum population density in a periodically varying environment 
relative to the static case is responsible for sustaining species coexistence in an extended region of parameter 
space.
Moreover, the extinction transition at high predation rate is moved to larger values of $\lambda / \sigma$ for
the simulation runs with periodically varying carrying capacities compared to systems with fixed environment.

\begin{figure}
    \includegraphics[width=\columnwidth]{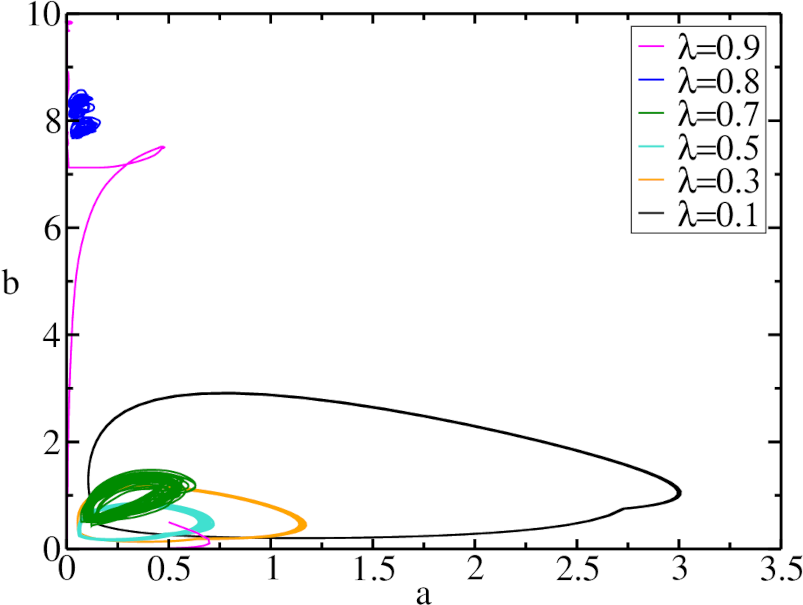}
    \caption{Predator-prey density phase space plots for various values of the predation rate $\lambda$ (as
    indicated), with $L=256$, $\sigma = \mu = 0.1$, $K_- = 1$, $K_+ = 10$, and $T_k = 100$. 
    The initial behavior of the system was discarded for all $\lambda$ values, except for $\lambda = 0.9$, for 
    which the predator population becomes extinct.}
    \label{fig:stochastic_phasespace}
\end{figure}

The predator-prey density phase space plots are constructed in Fig.~\ref{fig:stochastic_phasespace} by 
simulating the system for multiple predation rate values. 
We see that for each $\lambda$ the system fluctuates around a closed orbit. 
Upon increasing the predation rate $\lambda$, the radius of this closed orbit becomes smaller, while the
influence of stochastic fluctuations become more apparent. 
For $\lambda=0.8$, the orbit approaches $\rho_A=0$ which means that the predator population is close to 
extinction. 
Raising the predation rate further to $\lambda = 0.9$, the system reaches the (finite system size) absorbing 
state with vanishing predator density, see Fig.~\ref{fig:better_coexistence}.

\subsection{Fast and slow switching regimes}
\label{subsection:fast_slow_swtiching}

\begin{figure}
\subfloat[\label{subfig:stochastic_varying_period_a}$T_k=10$]{%
  \includegraphics[width=0.5\columnwidth]{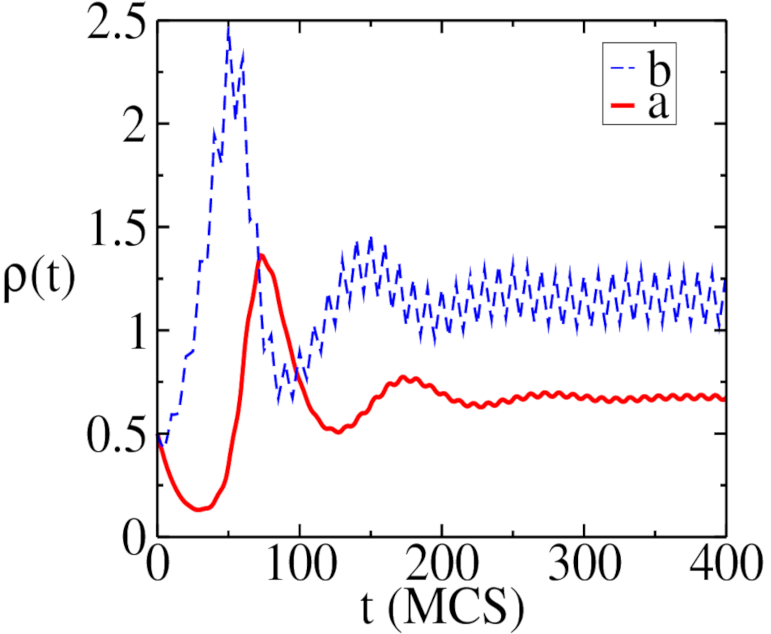}%
}\hfill
\subfloat[\label{subfig:stochastic_varying_period_b}$T_k=460$]{%
  \includegraphics[width=0.5\columnwidth]{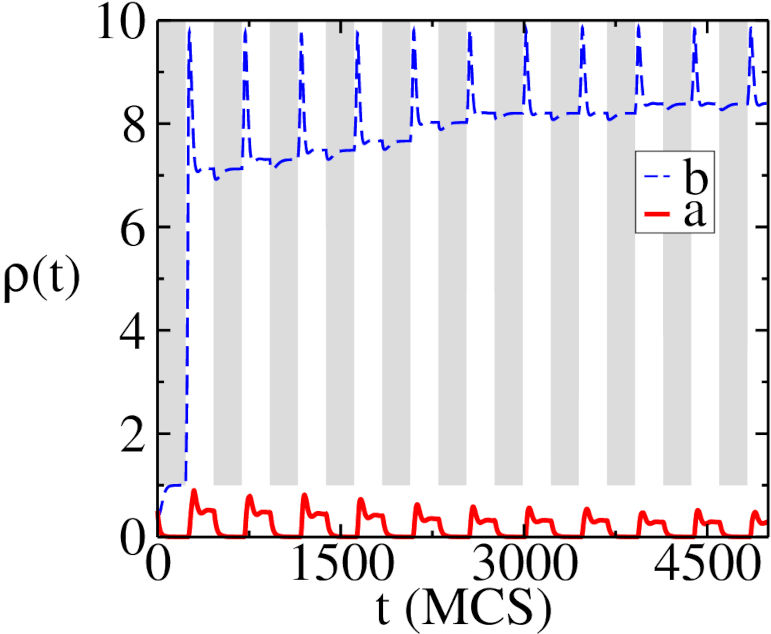}%
}
\caption{Predator (solid red) and prey (dashed blue) population densities averaged over $50$ realizations, 
for $L = 256$, $\sigma = \mu = \lambda = 0.1$, $K_- = 1$, $K_+ = 10$, and switching periods 
(a) $T_k = 10$, (b) $T_k = 460$, with the gray areas here indicating the population densities excluded by 
$K(t)$.}
\label{fig:stochastic_varying_period}
\end{figure}

We next carefully investigate how the system behaves in the two opposite limits of fast and slow 
environmental switching, relative to the intrinsic period of the Lotka--Volterra population oscillations. 
Figures~\ref{fig:stochastic_varying_period}(a) and (b) show the both populations' densities for $T_k = 10$
(fast switching) and $T_k = 460$ (slow switching). 
The time-averaged behavior of the density in the fast-switching regime resembles a system with a constant 
effective equivalent carrying capacity $K^*$ that should be related to $K_-$ and $K_+$. 
In the slow-switching regime the system is given sufficient time to approach a (quasi-)stationary state when 
$K_+ = 10$. 
The prey density then reaches very high values, and the system is slowly driven to predator extinction; 
however, it would take many cycles of the changing environment for this absorbing state to be attained. 
As the switching period $T_k$ is increased, the predator population may only survive for a few cycles; 
eventually, when $T_k$ is set too large, it will go extinct before the prey food resources become
abundant again. 

We now explore the equivalent static environment hypothesis in the fast-switching regime in more detail.
The mean-field rate equations suggest that for very short periods this equivalent carrying capacity equals the 
harmonic average ${\bar K}$ of $K_+$ and $K_-$, since Eqs.~(\ref{eq:mean-field}) only explicitly depend 
on $1 / K$. 
For longer periods, the mean-field model predicts that the dynamics becomes effectively equivalent to a 
quasi-static system with a rate-dependent equivalent carrying capacity $K^*$, Eq.~(\ref{eq:equivalent_K}).
One should expect the slow-switching equivalent carrying capacity in the stochastic lattice model to display a 
similar dependence on the microscopic reaction probabilities.
As mentioned earlier, their precise relationship with macroscopic reaction rates such as $\sigma$ and 
$\lambda$ is however subtle and difficult to capture quantitatively, which poses a problem for stringently 
testing Eq.~(\ref{eq:equivalent_K}) for the stochastic lattice model.
Yet for large $K_-$ and $K_+$, $K^*$ reduces approximately to the harmonic average ${\bar K}$, 
independent of the reaction rates. 
Hence we focus on testing the equivalent static environment hypothesis mainly with this effective carrying
capacity.

\begin{figure}[t]
\subfloat[\label{subfig:stochastic_K_eq_K_low_a}predator density]{%
  \includegraphics[width=0.5\columnwidth]{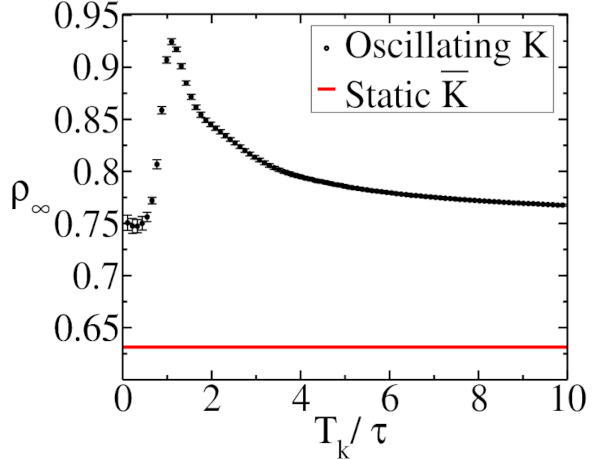}%
}\hfill
\subfloat[\label{subfig:stochastic_K_eq_K_low_b}prey density]{%
  \includegraphics[width=0.5\columnwidth]{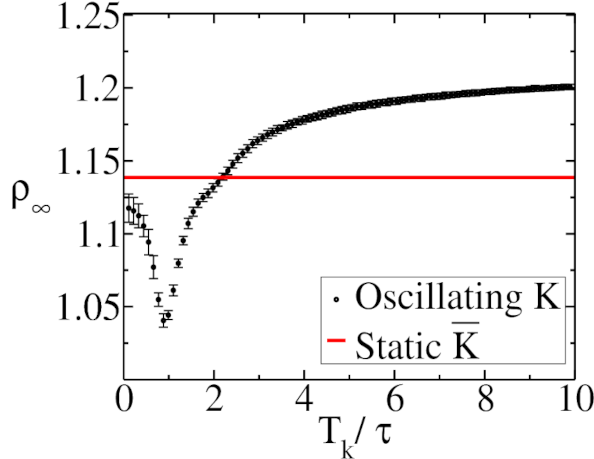}%
}
\caption{Long-time population densities $\rho_\infty$ averaged over six periods of the carrying capacity 
$K(t)$ plotted versus $T_k / \tau$, where $\tau$ denotes the intrinsic period of the equivalent static system 
with $K = \bar{K}$, where $L = 256$, $\sigma = \mu = \lambda = 0.1$, and $K_- = 2$, $K_+ = 6$ for the 
oscillating environment (black crosses), while $K = {\bar K} = 3$ for the static environment with fixed 
carrying capacity (full red).}
\label{fig:stochastic_K_eq_K_low}
\end{figure}
To this end, we first present Monte Carlo simulation data for our system with $K_-=2$ and $K_+=6$, hence 
${\bar K} = 3$, obtained for a series of different switching periods $T_k$, measured relative to the intrinsic
population oscillation period $\tau$ at fixed $\bar{K}$.
For comparison, we also display simulations with fixed carrying capacity $K = 3$, and display the resulting
population densities in Fig.~\ref{fig:stochastic_K_eq_K_low}.
We find that the predator density in the oscillating environment does not behave as if the environment were 
static with a harmonically averaged carrying capacity ${\bar K}$, with a discrepancy in the predator density of
at least $18.4\%$. 
In contrast, the time-averaged prey density $\rho_\infty$ matches with the static equivalent ${\bar K}$ value
for $T_k \approx 2.2 \tau$. 
Yet for faster switching rates, we observe worse agreement with a discrepancy of up to $5.45\%$. 
For periods $T_k > 2.2 \tau$, $\rho_\infty$ increases monotonically with $T_k / \tau$, deviating further from
the average prey density for the static equivalent ${\bar K}$. 
For larger $T_k / \tau$, the discrepancy between the harmonically averaged and the oscillating environments 
becomes more enhanced, although deviations remain less than $10\%$. 
Hence we conclude that our prey density data for an oscillating environment can be satisfactorily described  
by an equivalent constant environment for a wide range of oscillation periods. 
We note that both time-averaged population densities exhibit resonance-like extrema at $T_k \approx \tau$, 
owing to the environment switching just after the predators and prey have reached their maximum and 
minimum population counts, respectively, following their intrinsic Lotka--Volterra oscillations. 
As the period of the environment increases, more of these population oscillations may occur before the 
carrying capacity is reset, and integrating over one cycle of the environmental switching effectively averages 
over multiple periods of the intrinsic oscillations.
\begin{figure}[t]
\subfloat[\label{subfig:stochastic_K_eq_K_high_a}predator density]{%
  \includegraphics[width=0.5\columnwidth]{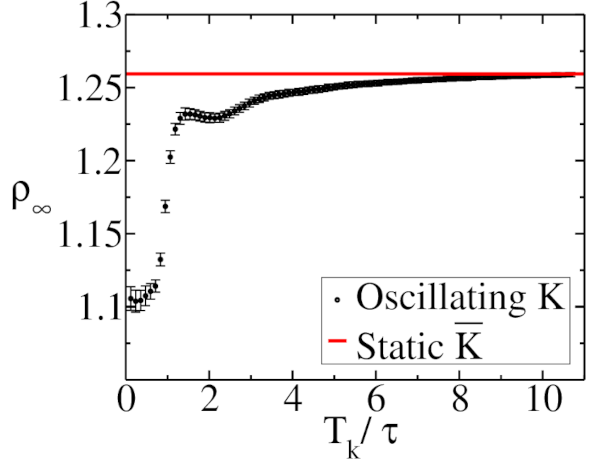}%
  }\hfill
\subfloat[\label{subfig:stochastic_K_eq_K_high_b}prey density]{%
  \includegraphics[width=0.5\columnwidth]{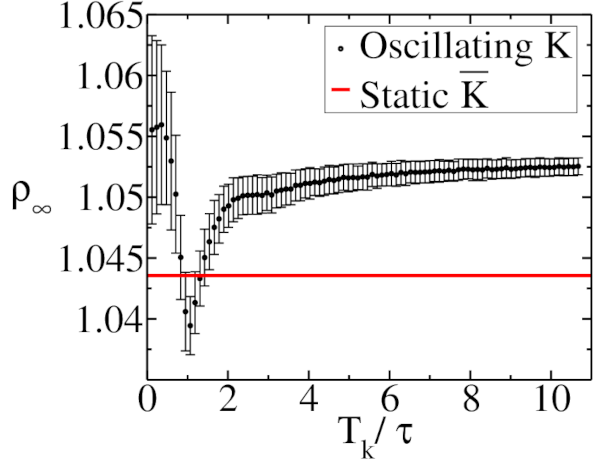}%
  }
\caption{Long-time population densities $\rho_\infty$ averaged over six periods of the carrying capacity 
$K(t)$ plotted versus $T_k / \tau$, where $\tau$ denotes the intrinsic period of the equivalent static system 
with $K = \bar{K}$, where $L = 256$, $\sigma = \mu = \lambda = 0.1$, and $K_- = 4$, $K_+ = 12$ for the 
oscillating environment (black crosses), while $K = {\bar K} = 6$ for the static environment with fixed 
carrying capacity (full red).}
\label{fig:stochastic_K_eq_K_high}
\end{figure}

\begin{figure*}[t]
\subfloat[\label{subfig:correlation_low_period_movie_a}$t=23$]{%
  \includegraphics[width=0.5\columnwidth]{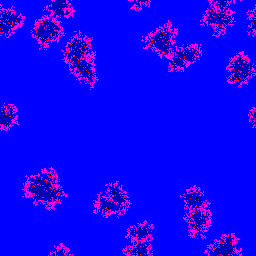}%
}\hfill
\subfloat[\label{subfig:correlation_low_period_movie_b}$t=28$]{%
  \includegraphics[width=0.5\columnwidth]{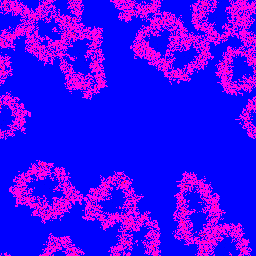}%
}\hfill
\subfloat[\label{subfig:correlation_low_period_movie_c}$t=50$]{%
  \includegraphics[width=0.5\columnwidth]{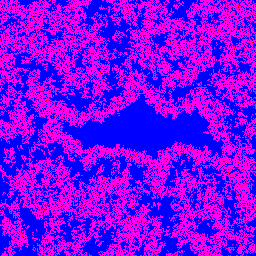}%
}\hfill
\subfloat[\label{subfig:correlation_low_period_movie_d}$t=101$]{%
  \includegraphics[width=0.5\columnwidth]{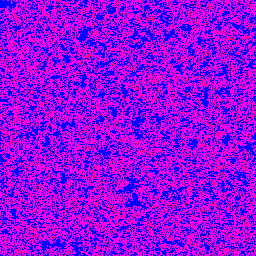}%
}
\caption{Snapshots of a single run for a system with parameters $L = 256$, $\sigma = \mu = 0.5$, 
$\lambda = 0.1$, $K_- = 1$, $K_+ = 10$, and $T_k = 10$; time $t$ is measured in units of Monte Carlo steps. 
The red and blue colored pixels indicate the presence of predators and prey, respectively, with the 
brightness representing the local density, the pink colored pixels pertain to sites with both predator and prey
present, and the black pixels represent empty sites. 
The system is initialized with $K(t=0) = K_- = 1$. The full movie can be viewed at \cite{note:movies}}
\label{fig:correlation_low_period_movie}
\end{figure*}

In Fig.~\ref{fig:stochastic_K_eq_K_high} we repeat this numerical investigation for $K_- = 4$, $K_+ = 12$, 
thus ${\bar K} = 6$. 
The time-averaged prey density $\rho_\infty$ for the oscillating environment agrees well
with the corresponding value for the static equivalent environment for all switching periods. 
However, the predator density for low periods does not match the harmonic mean hypothesis. 
For periods $T_k > \tau$, we see that the predator density with the oscillating environment approaches  
$\rho_\infty$ for the static equivalent environment. 
This suggests that the harmonically averaged carrying capacity works well to describe the mean predator
population density for large $K_-$ and $K_+$ values, and for large environment oscillation periods, such that 
the system reaches the stationary state before switching occurs. 
In conclusion, stochastic fluctuations may change the form of the general equivalent static carrying capacity 
(\ref{eq:equivalent_K}), yet it can still be approximated by the harmonic average for large carrying capacities.

Our simulation results indicate that the functional dependence of the prey density on the carrying capacity can 
be well approximated as $b \sim 1 / K$ for a large range of environmental switching periods. 
However, the predator density exhibits a more complicated dependence on the carrying capacity values and 
$T_k$; it can only be approximated by $a \sim 1 / K$ for large $K_-$ and $K_+$ and for $T_k \gg \tau$.
In the latter limit, the system reaches its quasi-stationary state before the environment switches, which for the 
used parameter values corresponds to a stable node with non-oscillatory kinetics; consequently, there is 
little variation with $T_k$.
Generally we observe that the long-time behavior of both population densities depends on the carrying 
capacity period in a non-monotonic manner. 

\subsection{Correlation functions}
\label{subsection:correlations}

The predator-prey pursuit and evasion waves characteristic of the stochastic spatial Lotka--Volterra model 
are more prominent in systems with high reaction rates. 
Therefore, we study the ensuing correlations for $\sigma = \mu = 0.5$, $\lambda = 0.1$, and leave 
$K_- = 1$, $K_+ = 10$. 
For these parameters the system resides deep in the predator extinction absorbing phase when $K = K_-$, 
and in the active two-species coexistence phase for $K = K_+$. 
The behavior of the system for environmental switching period $T_k = 10$ is exemplified by the simulation 
snapshots depicted in Fig.~\ref{fig:correlation_low_period_movie}. 
The predators are initially almost driven to extinction, but due to the switching environment the prey 
population increases until it fills most of the lattice. 
We observe that at $t = 23$ there remain only a few surviving predators which become localized sources for 
spreading waves. 
At $t = 28$, the prey may proliferate in the interior of the fronts as well, causing the population waves to 
spread both outwards and inwards, until they eventually collide and interfere with each other as seen at 
$t = 50$. 
Starting from $t = 101$, the lattice exhibits a global density oscillation, and it becomes difficult to discern the 
original locations of the wavefront sources.

\begin{figure}[b]
\subfloat[\label{subfig:correlations_low_period_a}auto-correlations]{%
  \includegraphics[width=0.5\columnwidth]{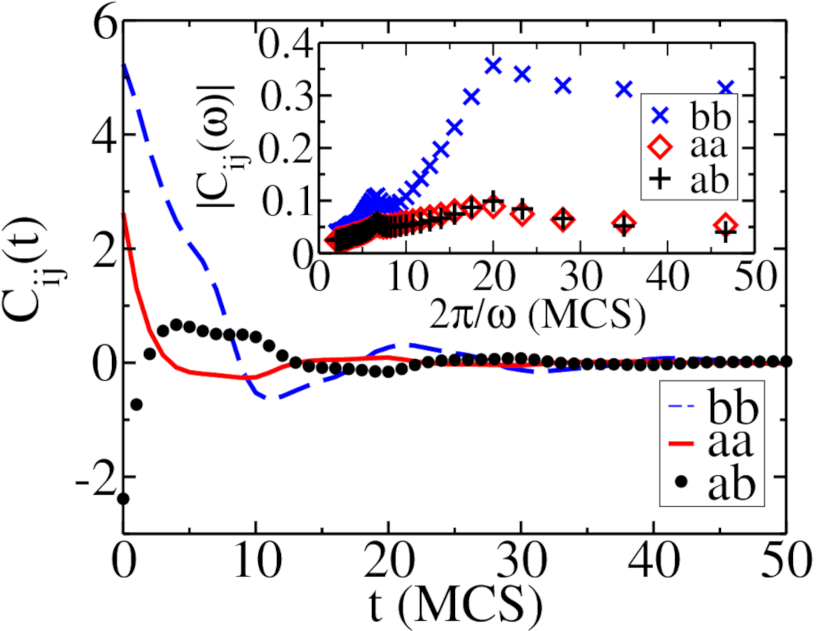}%
  }\hfill
\subfloat[\label{subfig:correlations_low_period_b}static correlations]{%
  \includegraphics[width=0.5\columnwidth]{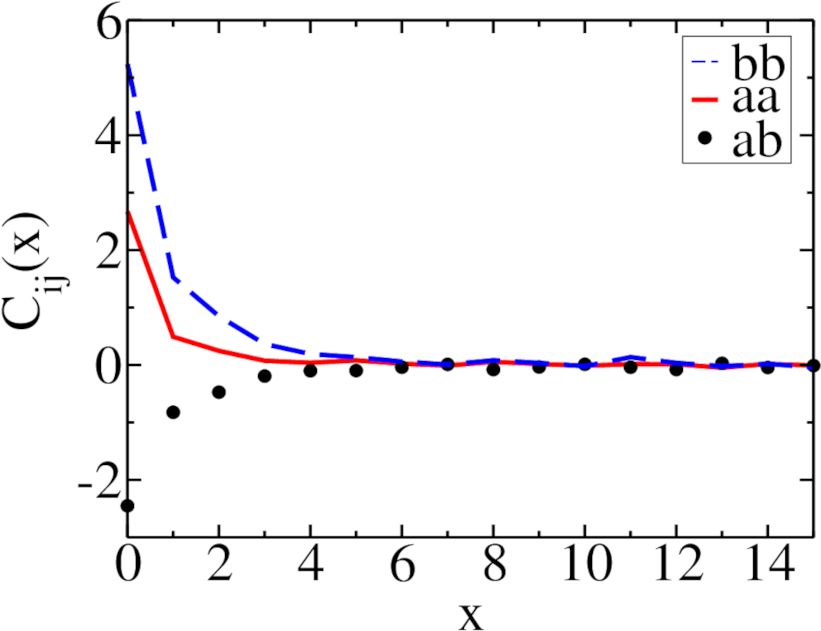}%
  }
\caption{Long-time correlation functions computed for a system with the following parameters: 
$L = 512$, $\sigma = \mu = 0.5$, $\lambda = 0.1$, $K_- = 1$, $K_+ = 10$, and $T_k = 10$. 
(a) Temporal auto-correlations computed for $t_0 = 1000$, with $t$ measured starting from $t_0$. 
The inset shows the Fourier transform of the auto-correlation time series. 
This data was averaged over $10,000$ ensembles and for $512$ lattice sites, giving an equivalent of a total of 
$5,120,000$ independent ensembles.
 (b) Static correlation functions taken at $t_0 = 1000$. 
Distances $x$ are measured in units of the (dimensionsless) lattice spacing; data averaged over $10,000$ 
distinct ensembles.}
\label{fig:correlations_low_period}
\end{figure}

\begin{figure*}[t]
\subfloat[\label{subfig:correlations_high_period_movie_a}$t=39$]{%
  \includegraphics[width=0.5\columnwidth]{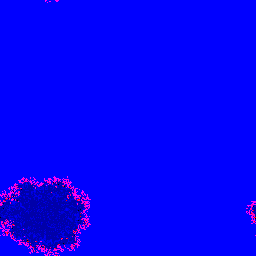}%
}\hfill
\subfloat[\label{subfig:correlations_high_period_movie_b}$t=56$]{%
  \includegraphics[width=0.5\columnwidth]{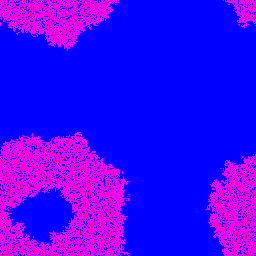}%
}\hfill
\subfloat[\label{subfig:correlations_high_period_movie_c}$t=72$]{%
  \includegraphics[width=0.5\columnwidth]{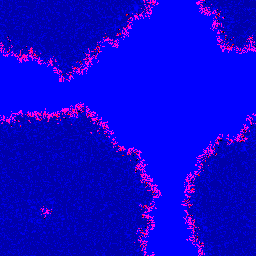}%
}\hfill
\subfloat[\label{subfig:correlations_high_period_movie_d}$t=264$]{%
  \includegraphics[width=0.5\columnwidth]{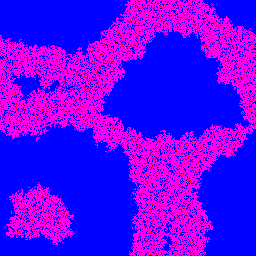}%
}
\caption{Snapshots of a single run for a system with parameters $L = 256$, $\sigma = \mu = 0.5$, 
$\lambda = 0.1$, $K_- = 1$, $K_+ = 10$, and $T_k = 30$; time $t$ is measured in units of Monte Carlo steps. 
The red and blue colored pixels indicate the presence of predators and prey, respectively, with the 
brightness representing the local density, the pink colored pixels pertain to sites with both predator and prey
present, and the black pixels represent empty sites. 
The system is initialized with $K(t=0) = K_- = 1$. The full movie can be viewed at \cite{note:movies}}
\label{fig:correlations_high_period_movie}
\end{figure*}
\begin{figure*}
\subfloat[\label{subfig:correlations_high_period_a}auto-correlations]{%
  \includegraphics[width=0.5\columnwidth]{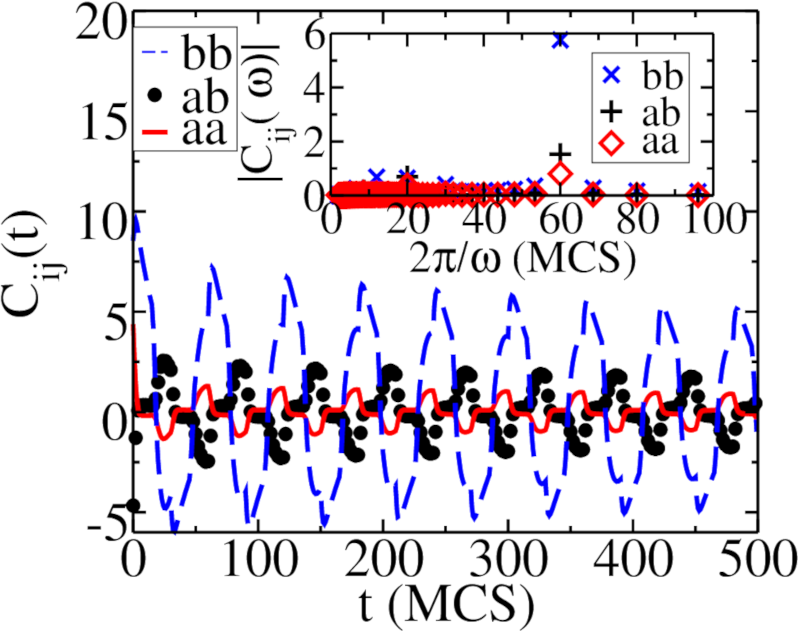}%
}\hfill
\subfloat[\label{subfig:correlations_high_period_b}$ij=aa$]{%
  \includegraphics[width=0.5\columnwidth]{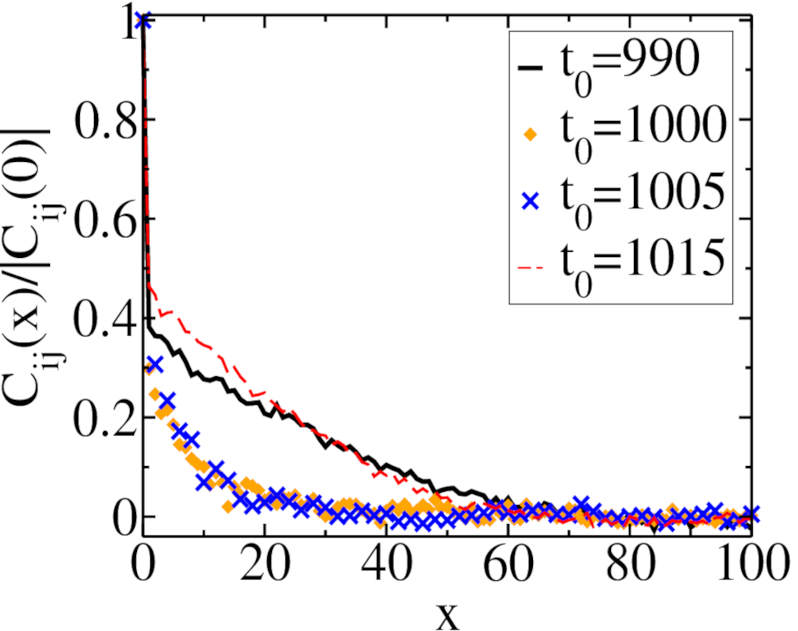}%
}\hfill
\subfloat[\label{subfig:correlations_high_period_c}$ij=ab$]{%
  \includegraphics[width=0.5\columnwidth]{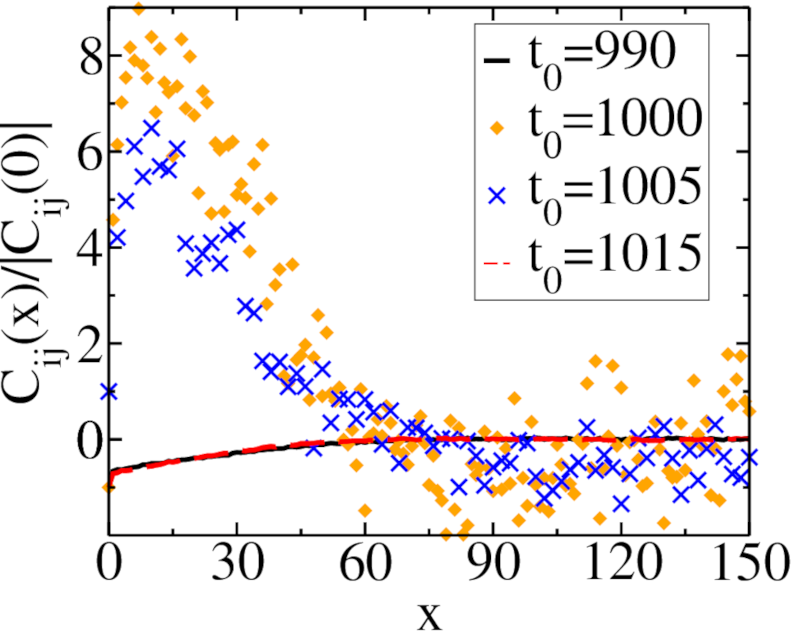}%
}\hfill
\subfloat[\label{subfig:correlations_high_period_d}$ij=bb$]{%
  \includegraphics[width=0.5\columnwidth]{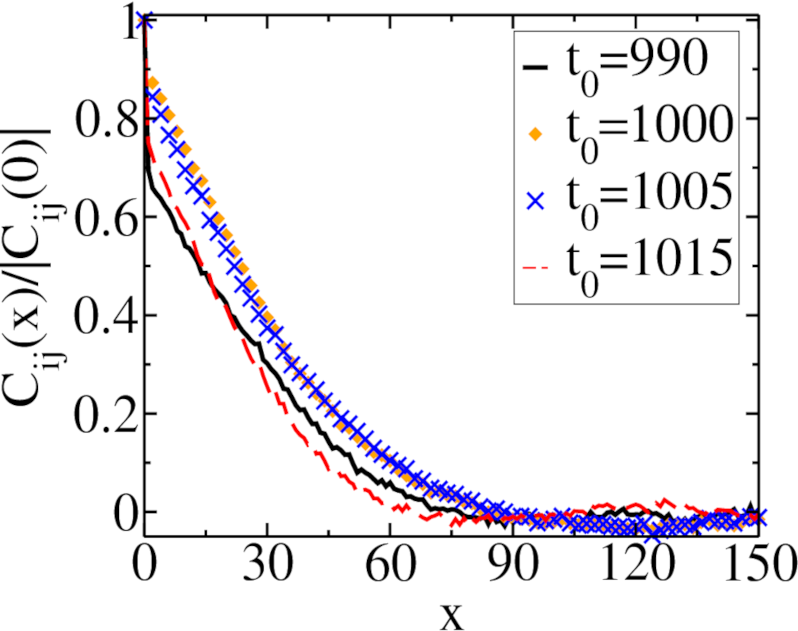}%
}
\caption{Long-time correlation functions computed for a system with the following parameters: 
$L = 512$, $\sigma = \mu = 0.5$, $\lambda = 0.1$, $K_- = 1$, $K_+ = 10$, and $T_k = 30$. 
(a) Temporal auto-correlations computed for $t_0 = 990$, with $t$ measured starting from $t_0$. 
The inset shows the Fourier transform of the auto-correlation time series. 
(b) Static predator-predator,  (c) predator-prey, and (d) prey-prey correlation functions for different values of 
$t_0$, normalized by $|C_{ij}(x=0)|$; distances $x$ are measured in units of the lattice spacing. 
Data averaged over $10,000$ distinct ensembles.}
\label{fig:correlations_high_period}
\end{figure*}
The associated temporal auto- and static correlation functions are displayed in 
Fig.~\ref{fig:correlations_low_period}. 
The auto-correlation functions exhibit damped oscillations with a peak period $2 T_k = 20$, twice the 
switching period of the carrying capacity. 
This is due to the fact that the two-point correlation function contains a product of particle densities, and the 
square of sinoidal functions may be decomposed into sine functions with doubled period. 
Note that the auto-correlations decay to zero after approximately $40$ time steps. 
The on-site population restrictions induce anti-correlations between individuals of the same species; the 
cross-correlation function $C_{ab}$ becomes positive after some time has elapsed., indicating that surviving
predators follow the prey with some time delay. 
The static correlation functions rapidly decay to zero, demonstrating that the spatial correlation lengths are 
small, on the scale of a few lattice spacings.

Figure~\ref{fig:correlations_high_period_movie} shows simulation snapshots for the system parameters,
but with a larger switching period $T_k = 30$. 
In this run, only one predator patch has survived by $t = 39$. 
Subsequently it serves as a source for a spreading population wave that later interferes with itself owing to 
the periodic boundary conditions of the lattice. 
At $t = 56$ the wave starts spreading in both directions until at $t = 72$, when the system returns to the 
low carrying capacity regime, and the prey in the interior of the front are not allowed to reproduce further. 
Even after a long time period at $t = 264$, there is only a single density oscillation center that is sourced by 
the sole predator patch that had survived at $t = 39$.

In Fig.~\ref{subfig:correlations_high_period_a} we plot the corresponding auto-correlation functions, which 
exhibit a much slower decay compared to Fig.~\ref{fig:correlations_low_period}(a) for $T_k = 10$. 
This suggests that a carrying capacity period of $T_k=30$ causes a resonance effect, which indeed becomes
apparent in the simulation movies, as in this case the switching happens approximately when the waves travel 
back to the location of the source. 
The resonance sustains the spatial and temporal correlations and thereby stabilizes the travelling waves, 
leading to a sustained asynchrony that promotes species coexistence.
The Fourier transform again confirms that the auto-correlation functions oscillate with a period $2 T_k$.
Since the carrying capacity switching period is $T_k = 30$, and it is initialized with $K(t=0) = K_-$, the
behavior of the system at different $t_0$ values can be described as follows:
For $t_0 = 990$, the system has just switched from $K(t) = K_+$ to $K_-$; at $t_0 = 1000$, it still resides 
at carrying capacity $K_-$; for $t_0  =1005$, the system has just switched from $K_-$ back to $K_+$; at
$t_0 = 1015$, the carrying capacity is still $K_+$.
The static correlation functions, shown in Figs.~\ref{fig:correlations_high_period}(b,c,d), exhibit similar 
behavior for $t_0 = 990$ and $t_0 = 1015$, and for $t_0 = 1000$ and $t_0 = 1005$, respectively, which 
suggests a common delay time for the correlations. 
At $t_0 = 990$, the system is in the state with $K(t) = K_-$, while at $t_0 = 1015$, $K(t)  =K_+$, about to 
switch to $K_-$; and similarly at $t_0 = 1005$ and $t_0 = 1000$.
Compared with the system with faster switching period $T_k = 10$, the static correlations decay over a larger 
distance, in agreement with the movies and snapshots which show wider wavefronts. 
The predator-prey cross-correlation function $C_{ab}(x)$ displays maxima at positive values for 
$t_0 = 1000$ and $t_0 = 1005$, when the carrying capacity is low and few individuals are present per site. 
Conversely at $t_0 = 990$ and $t_0 = 1015$, when the population densities are large, the only positive peak 
occurs at $x = 0$, due to the fact that predators tend to be on the same site as prey for large $K$. 
For low carrying capacities, the predators cannot reside on the same locations as the prey, so instead they are 
most likely to be in the close prey neighborhood.

\subsection{Asymmetric switching intervals}
\label{subsection:asymmetric}

\begin{figure}
\subfloat[\label{subfig:asymmetric_a}density time-series]{%
  \includegraphics[width=0.5\columnwidth]{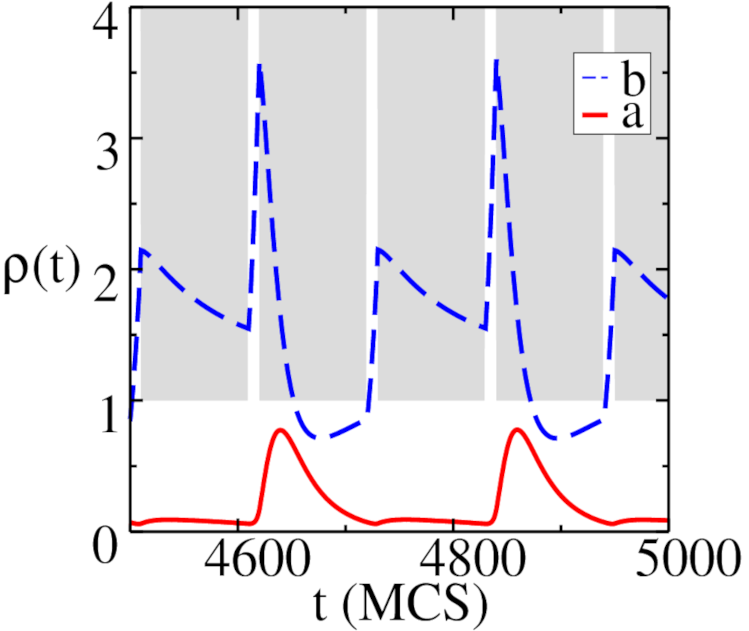}%
  }\hfill
\subfloat[\label{subfig:asymmetric_b}Fourier transform]{%
  \includegraphics[width=0.5\columnwidth]{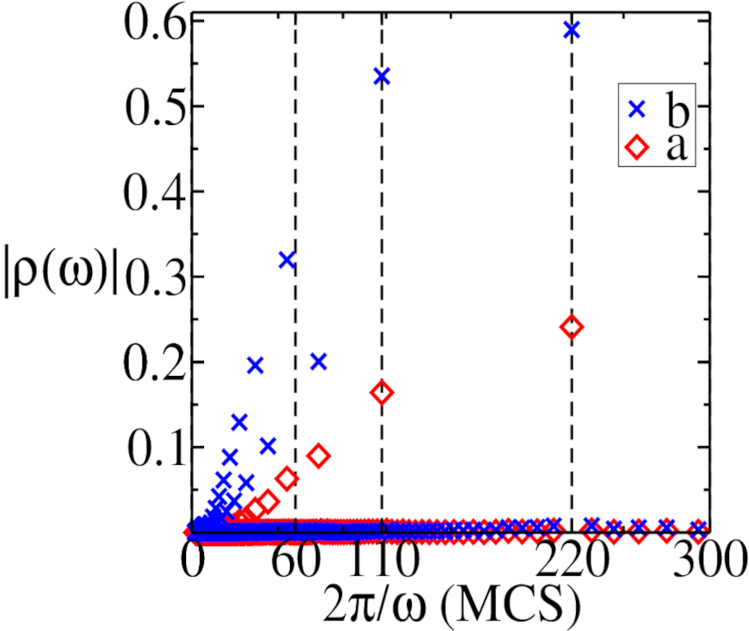}%
  }
\caption{(a) Predator and prey densities averaged over $50$ realizations for a system with asymmetric switching 
intervals: $L = 256$, $\sigma = \mu = \lambda = 0.1$, $K_- = 1$, $K_+ = 10$, $T_- = 100$, 
$T_+ = 10 = 0.1 \, T_-$; the shaded gray areas are excluded by the switching carrying capacity $K(t)$. 
(b) Fourier transforms of the population density time evolution in (a).}
\label{fig:asymmetric}
\end{figure}

Finally, we further investigate the properties of our system by applying an asymmetric square signal for the
switching carrying capacity, such that $K = K_-$ for $T_-$ time steps, and then $K = K_+$ for the subsequent
time interval of length $T_+$, where $T_- \neq T_+$. 
The total switching period of the carrying capacity then is $T_k = T_- + T_+$. 
Simulating such a stochastic lattice system reveals a period-doubling effect for an intermediate range of 
$T_+ / T_-$ ratios, as shown in Fig~\ref{fig:asymmetric}. 
For either too small or too large time interval ratios, no period-doubling effect could be observed. 
The origin of this intriguing period-doubling effect appears to be that prey particles are not able to reproduce 
quickly enough while the system has attained the high carrying capacity $K_+$. 
Hence, it takes the system two cycles of the oscillating environment for the prey density to reach its peak value.

\section{Conclusion and outlook}
\label{section:conclusion}

In this paper, we have investigated the paradigmatic Lotka--Volterra predator-prey model with a periodically 
varying carrying capacity $K(t)$ that represents seasonally changing food resource availability for the prey 
population. 
The model was studied both by a mean-field analysis based on the deterministic rate equations, and through
detailed individual-based stochastic Monte Carlo simulations on a two-dimensional lattice with periodic boundary
conditions.
Both the mean-field and the stochastic lattice model exhibit characteristic periodic behavior induced by the 
changing environment.
The rate equation solutions display a region in parameter space with period-doubling and period-quadrupling 
features; such effects are naturally expected in driven non-linear dynamical systems. 
However, the period-doubling region in parameter space is not observed in the stochastic lattice model: 
The internal stochastic noise evidently dominates and eliminates these non-linear effects.
Yet we were able to induce period-doubling dynamics in the lattice model by utilizing an external periodic drive
signal with asymmetric switching intervals.

The phase space analysis demonstrated that, for parameters that lead to an ecologically stable system (which 
does not evolve into an absorbing population exctintion state), the phase space orbits are closed loops, whose
sizes decrease with growing predation rate $\lambda$, indicating that the population oscillation amplitudes
become reduced with enhanced predation efficiency.
A periodically varying environment allows the system to remain stable even for lower values of $\lambda$, as
compared to the corresponding system with fixed carrying capacity. 
We find that the periodically varying environment induces oscillations with greater amplitudes, without hitting 
the predator extinction absorbing state. 
We argue that this phenomenon is due to the density oscillations extending beyond their 
maximum static values when periodically switching between low and high carrying capacity environments. 
Hence scarcity of food resources in one season induces a higher species density (relative to a constant 
environment) in later seasons when food resources become more abundant again.
Furthermore, we observe that even for the same value of $\lambda$, the periodically varying systems display
larger oscillation amplitudes than the static system. 
The finite system size extinction threshold at high predation rates is shifted to higher values of 
$\lambda$ as well.
Thus, a periodically changing, externally driven environment leads to a richer ecology and promotes species 
diversity.

We investigated the long-time behavior of the population densities by studying their averages over multiple 
cycles of the periodic environment as a function of switching period $T_k$. 
For the mean-field model, the prey density average does not depend on $T_k$, and is equal to its 
$K$-independent stationary value. 
In contrast, the mean predator density turns out equal to the stationary value of a static equivalent $K^*$
value given by Eq.~(\ref{eq:equivalent_K}) that for small periods simply reduces to the harmonic average 
${\bar K}$ of $K_-$ and $K_+$.
Interestingly, for intermediate periods $T_k$ one encounters a non-monotonic crossover regime between 
these two averages for intermediate values of the period with characteristic resonant features when $T_k$ is
close to the intrinsic Lotka--Volterra population oscillation period. 
The stochastic lattice model reveals more complex behavior owing to renormalization of the equivalent
stationary carrying capacity values as well as the reaction rates. 
The mean stationary prey density value is no longer $K$-independent, and it shows non-monotonic behavior 
as a function of $T_k$. 
Nevertheless, quantitatively these effects are small, implying that the harmonically averaged equivalent 
stationary value ${\bar K}$ gives a good approximation for the long-time average of the prey density. 
The predator density average matches the stationary equivalent ${\bar K}$ only for high values of $K_-$ and 
$K_+$, as well as large switching periods $T_k$. 

We evaluated the auto-correlation and static correlation functions for the stochstic lattice model specifically 
for two different periods, $T_k = 10$ and $T_k = 30$. 
The Fourier transformed auto-correlations exhibit peaks at $2 T_k$. 
Due to the local on-site restrictions, the cross-correlation functions are negative at short distances. 
For the smaller period $T_k = 10$, the auto-correlations decay to zero already after about a single oscillation 
period, and the static correlations rapidly decay to zero as well, indicating a small spatial correlation length. 
When the period is increased to $T_k = 30$, we observe a resonance effect causing the auto-correlations to 
decay at a much slower rate. 
As the simulations movies show, this resonant behavior is caused by the spherical travelling activity waves 
pulsing back to the location of their sources.
For low $T_k$, the interference of population waves with each other seems to average
out local structures, and instead lead to a global temporal oscillation with the external frequency prescribed
by the carrying capacity switches. 
Consequently, prolonged spatial correlations are not observed for $T_k=10$, in contrast with the data for
longer $T_k=30$.
The static correlation functions for $T_k=30$ exhibit a much slower decay as well, indicating markedly 
longer-ranged correlations. 
Plotting the static correlation functions at different times, we detect a time-delay effect, where the stationary 
correlations require some time to respond to the changing environment.
This is in contrast to the population densities (one-point functions) which respond almost
instantaneously to the switching environment.

Using our observations pertaining to the long-time behavior of the population densities in the mean-field 
model, we obtained a closed-form solution that approximates the quasi-stationary state of the system for a 
fast switching carrying capacity; more preciely, this solution holds if $T_k \sqrt{\lambda |db / dt|} \ll 1$. 
We were able to explicitly demonstrate the regime of applicability for this approximation, 
c.f.~Fig.~\ref{fig:error_plot}. 
It should be possible to utilize this asymptotic technique to study generalizations to other periodically 
varying variables, e.g., varying reaction rates, to shed light on the response of such systems to sudden 
parametric variations.

The importance of the environment on the balance of animal populations has been understood at least 
since the work of Nicholson \cite{varyenv-ref3}. 
Spatial models offer rich behavior due to an enhancement in species coexistence \cite{spatial1, spatial2, 
spatial3, spatial4}, and seasonally varying environments are known to promote species coexistence even 
further \cite{seasonal1,seasonal2,seasonal3}. 
In this numerical and analytical study of the effects of a periodically varying carrying capacity, we determined
the shift in population balance in both fast- and slow-switching limits, by showing that the system can then 
be described via oscillations around a quasi-fixed point. 
In the crossover regime, the species balance depends on the environmental modulation period. 
We utilized a stochastic lattice model to investigate how resonance affects the pursuit and evasion waves, 
which are known to enhance species coexistence through the asynchrony effect described in 
Refs.~\cite{async1,spatial2,spatial4}. 
Our results show that seasonal changes at resonance stabilize the intrinsic dynamic correlations of the system
that in turn support asynchronous states, thus enabling predators to survive even if they are at a severe 
disadvantage during one of the seasons. 
The sustainability of predators could also be attributed to the growth rate at low density described in 
Ref.~\cite{envxx2}:
As the environment switches from low to high carrying capacity, this growth rate suddenly jumps to a high 
value, which is responsible for maintaining the predator population.

The description of reaction-diffusion systems in terms mean-field rate equations is of course useful, and 
often provides an accurate qualitative description of real systems for some region in parameter space. 
However, this paper demonstrates that when an ecological system is subjected to periodic variations in the 
environment, a proper stochastic model may behave differently than its mean-field representation. 
Fluctuations can lead to dramatic changes in the behavior of the system as the present results indicate. 
One method of steering ecological communities towards a certain desirable behavior is to alter the 
environment. 
Therefore, developing succesful control schemes for such systems requires taking the effects of fluctuations 
into proper consideration. 
A full understanding of the fundamental problem of species diversity, and beyond, constructing a 
quantitative theory of biological evolution, hinge on unraveling the impact of environmental dynamics on 
ecological systems.

\begin{acknowledgments}
We would like to thank Matthew Asker, Marco Brizzolara, Jason Czak, Llu\'is Hernandez-Navarro, Hana Mir, 
Mauro Mobilia, Michel Pleimling, Alastair Rutlidge, James Stidham, Brian Thibodeau, Louie Hong Yao, and 
Canon Zeidan for their helpful feedback on our work. 
We are grateful to Rana Genedy for reading the manuscript draft and providing suggestions for improvement. 
This research was supported by the U.S National Science Foundation, Division of Mathematical Sciences under 
Award No. NSF DMS-2128587.
\end{acknowledgments}

\bibliographystyle{apsrev4-2}
\nocite{*}

\bibliography{mybib} 

\end{document}